\documentclass[reprint,onecolumn,notitlepage,longbibliography]{revtex4-1}

\usepackage{amsmath,amssymb}    
\usepackage{graphicx}   
\usepackage{verbatim}   
\usepackage[colorinlistoftodos]{todonotes}  

\usepackage{color}      

\usepackage{hyperref}   

\usepackage{bm}
\usepackage{upgreek} 
\usepackage{ulem}

\graphicspath{{./Figures/}}

\begin{document}
\title{Competing effects of inertia, sheet elasticity, and fluid viscoelasticity on the synchronization of two actuated sheets}
\author{Chaojie Mo}
\author{Dmitry A. Fedosov}\email{d.fedosov@fz-juelich.de}
\affiliation{Theoretical Physics of Living Matter, Institute of Biological Information Processing and Institute for Advanced Simulation,
Forschungszentrum J\"ulich, 52425 J\"ulich, Germany}
\date{\today}  

\begin{abstract}
Synchronization of two actuated sheets serves as a simple model for the interaction between flagellated microswimmers. Various 
factors, including inertia, sheet elasticity, and fluid viscoelasticity, have been suggested to facilitate the synchronization 
of two sheets; however, the importance of different contributions to this process still remains unclear. We perform a systematic
investigation of competing effects of inertia, sheet elasticity, and fluid viscoelasticity on the synchronization of two sheets. 
Characteristic time $\tau^\mathrm{s}$ for the synchronization caused by inertial effects is inversely proportional to sheet Reynolds 
number $\mathrm{Re}$, such that $\tau^\mathrm{s} \omega \propto \mathrm{Re}^{-1}$ with $\omega$ being the wave frequency. 
Synchronization toward stable in-phase 
or opposite-phase configuration of two sheets is determined by the competition of inertial effects, sheet elasticity,  
fluid compressibility and viscoelasticity. Interestingly, fluid viscoelasticity results in strong synchronization forces for 
large beating amplitudes and Deborah numbers $\mathrm{De} > 1$, which dominates over other factors and favors the in-phase 
configuration. Therefore, our results show that fluid viscoelasticity can dramatically enhance synchronization of microswimmers. 
Our investigation deciphers the importance of different competing effects for the synchronization of two actuated sheets, leading to 
a better understanding of interactions between microswimmers and their collective behavior.
\end{abstract}

\maketitle

\section{Introduction}

Locomotion of biological and artificial microswimmers and their collective behavior have attracted considerable scientific and 
technological attention recently \cite{Lauga_HSO_2009,Elgeti_PMS_2015,Bechinger_APC_2016,Palagi_BIR_2018}. The foci of such studies range 
from physical mechanisms governing the motion and interaction of microswimmers to their use in practical applications and 
the emergence of collective behavior. One of the interesting aspects is the interaction between multiple swimmers facilitated 
by a suspending fluid medium. For instance, swimming spermatozoa tend to synchronize their beating flagella when they are close to 
each other \cite{Woolley_SSF_2009,Yang_CS2D_2008,Nosrati_2DS_2015}. Even though distinct spermatozoa likely have differences in their intrinsic 
properties, they are able to adjust their beating characteristics (e.g., phase and frequency), and swim together as a concerted 
unit \cite{Woolley_SSF_2009}. Furthermore, synchronization of motion mediated by suspending medium is relevant for many other 
microswimmers, which propel using helical flagella \cite{Kim_MSM_2003,Reigh_SBF_2012} or cilia \cite{Brumley_HSW_2012,Elgeti_EMW_2013}. 

One of the first propositions that the synchronization of microswimmers is mediated by hydrodynamic interactions corresponds 
to the theoretical work of Taylor in 1951 \cite{Taylor_ASO_1951} for two waving tails. Interestingly, the first experimental 
confirmation of the importance of hydrodynamic interactions for the synchronization of two beating flagella has been realized 
only a few years ago \cite{Brumley_FSH_2014}. Theoretical analysis of microswimmer behavior and 
possible synchronization interactions is generally performed under the assumption of zero Reynolds number (i.e., no inertia)
\cite{Taylor_ASO_1951,Pedley_HPS_1992,Lauga_HSO_2009,Golestanian_HSR_2011}, because of non-linearity of the inertial term in 
the Navier-Stokes equations. Several theoretical studies have also considered the effect of fluid inertia at small Reynolds numbers 
\cite{Reynolds_SMO_1965,Tuck_NSP_1968}. Even though the assumption of vanishing inertia is generally justified by the small size and low swimming velocity of 
microswimmers, there are examples of artificial microrobots which operate at non-negligible Reynolds numbers \cite{Palagi_BIR_2018,Diller_CDM_2014,Huang_SMM_2016}.  
A theory, in which the synchronization of two inextensible waving sheets is 
considered, predicts no synchronization of the sheets having a front-back motion symmetry (e.g., a pure sine wave) due to 
kinematic reversibility of Stokes flow (i.e., under the assumption of no inertia) \cite{Elfring_HPL_2009}. Thus, synchronization
is only possible if there exist additional irreversible factors which break the symmetry \cite{Elfring_HPL_2009,Friedrich_HSO_2016}.
For example, to make the synchronization of two sheets possible, a front-back asymmetry in the beating motion is proposed
\cite{Elfring_HPL_2009,Elfring_PHS_2011}. Furthermore, other factors, such as non-negligible inertia
\cite{Fauci_IOF_1990,Fauci_SMB_1995,Theers_SRM_2013}, sheet elasticity \cite{Elfring_SFS_2011}, and viscoelasticity of 
non-Newtonian fluids \cite{Elfring_2DS_2010,Chrispell_AES_2013}, are sufficient to break the symmetry and enable synchronization. 

Another interesting aspect in the synchronization of two sheets is that there exist two stable synchronized configurations, 
namely in-phase and opposite-phase conformations with a phase difference $\phi_\mathrm{d}=0$ and $\phi_\mathrm{d}=\pi$ between 
the two sheets, respectively. Either the in-phase or opposite-phase configuration is stable, depending on various conditions. 
For instance, the geometry of a prescribed asymmetric wave determines the preference for each configuration
\cite{Elfring_HPL_2009,Elfring_PHS_2011}. In case of non-negligible inertial effects, the opposite-phase conformation is preferred 
with increasing Reynolds number \cite{Fauci_IOF_1990}. Sheet flexibility 
\cite{Elfring_SFS_2011} as well as viscoelasticity of an Oldroyd-B fluid \cite{Elfring_2DS_2010,Chrispell_AES_2013} drive 
the system toward the in-phase configuration. Noteworthy, the theory of sheet synchronization in Oldroyd-B fluids 
predicts the strongest synchronization force at Deborah number $\mathrm{De}$ of unity, while at large $\mathrm{De}$,
the synchronization forces asymptotically approach zero. For comparison, Deborah number of a swimming sperm in cervical 
mucus is larger than $100$, and fluid viscoelasticity dramatically enhances clustering of bovine sperm \cite{Tung_FVE_2017}.     
Despite several existing studies on sheet synchronization, the interplay and importance of different competing effects 
remains unclear.   

We perform a systematic analysis of the importance of different aforementioned factors for the synchronization of two sheets.
Our simulations are based on the smoothed dissipative particle dynamics (SDPD) method 
\cite{Espanol_SDPD_2003,Mueller_SDPD_2015,Quesada_VEF_2009}, a particle-based hydrodynamics technique, where both Newtonian and 
Oldroyd-B fluids are implemented. 
Two different setups are considered, including (i) a pair of inextensible waving sheets with a prescribed motion, for which 
synchronization forces are measured, and (ii) two flexible sheets with an internal actuation, for which dynamic synchronization 
toward one of the stable configurations is simulated. Our results show that for any non-zero Reynolds number $\mathrm{Re}$, the two sheets 
always synchronize regardless of its magnitude. When inertial effects dominate, the opposite-phase configuration is preferred and 
a characteristic time $\tau^\mathrm{s}$ for the synchronization normalized by the wave frequency $\omega$ is inversely proportional 
to $\mathrm{Re}$, i.e. $\tau^\mathrm{s} \omega \propto  \mathrm{Re}^{-1}$.

Sheet elasticity also affects stable synchronized configuration, driving the two sheets toward the in-phase configuration. 
Fluid viscoelasticity, when a dominating factor, also
drives the sheets toward the in-phase configuration. Nevertheless, at high enough $\mathrm{Re}$, inertial effects may favor the 
opposite-phase conformation even in viscoelastic fluids. The modes of stable synchronized configurations with respect to 
different factors are summarized in Table \ref{tab:sum}.  Furthermore, note that the failure of the theory in Ref.~\cite{Elfring_2DS_2010} 
to predict large synchronization forces at $\mathrm{De} > 1$ is related to the leading order approximation in terms of the 
wave amplitude. For large enough wave amplitudes, strong deviations in synchronization forces with respect to the theoretical 
predictions are observed for $\mathrm{De} > 1$, leading to a dramatic enhancement of sheet synchronization by fluid viscoelasticity. 
These results are consistent with experimental observations of the pronounced enhancement of sperm clustering in viscoelastic fluids
\cite{Tung_FVE_2017}. In conclusion, our results provide better understanding of different competing effects for sheet synchronization
and can be used to control the synchronization of artificial swimmers.

The paper is organized as follows. Section \ref{sec:model} presents fluid and sheet models, simulation setup, and the validation 
of these models against available theoretical predictions. In section \ref{sec:newton}, synchronization of two sheets in Newtonian 
fluids is studied for different model parameters affecting the value of $\mathrm{Re}$, fluid compressibility, and sheet flexural 
rigidity. Section \ref{sec:nonnewton} presents synchronization results in Oldroyd-B viscoelastic fluids.  Swimming 
efficiency of two synchronized sheets is discussed in section \ref{sec:eff}. Finally, we conclude in section \ref{sec:conc}.

\begin{figure}[!htb]
     \centering
     \includegraphics[width=.6\linewidth]{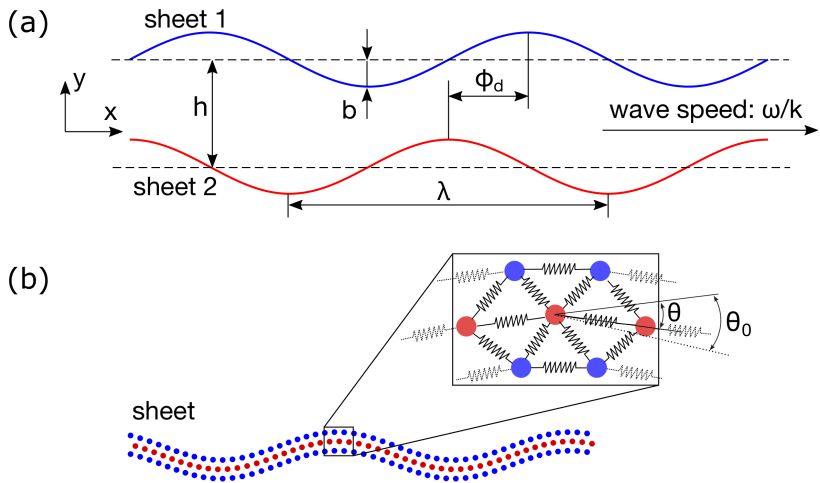}
     \caption{Model schematic. (a) Setup and basic parameters of the two actuated sheets. Here, $b$ is the wave amplitude, 
     $k$ is the wave number, $\lambda=2\pi / k$ is the wave length, $\omega$ is the wave frequency, such that 
     the wave speed is $\omega/k$. $h$ is the distance between average positions of the sheets and $\phi_\mathrm{d}$ is 
     a phase difference between their actuations. (b) Model representation of a flexible sheet constructed by three particle layers 
     interconnected by springs. $\theta$ is the instantaneous angle between two adjacent springs in the middle layer (marked in red), and 
     $\theta_0$ is the spontaneous angle employed for sheet actuation.}    
     \label{fig:01}
\end{figure}

\section{Methods and models}
\label{sec:model}

\subsection{Viscoelastic fluid model}

Fluid flow is modeled by the smoothed dissipative particle dynamics (SDPD) method \cite{Espanol_SDPD_2003,Mueller_SDPD_2015},
which is a particle-based mesoscopic hydrodynamics approach. SDPD is derived through a Lagrangian discretization of the Navier-Stokes 
equations similar to the smoothed particle hydrodynamics (SPH) method \cite{Monaghan_SPH_1992}, with the proper inclusion of thermal 
fluctuations following the dissipative particle dynamics (DPD) approach \cite{Hoogerbrugge_SMH_1992,Espanol_SMO_1995}. We employ 
an SDPD version which conserves angular momentum \cite{Mueller_SDPD_2015}, as it can be crucial for some problems
\cite{Hu_AMC_2006,Gotze_RAM_2007}. In SDPD, each particle can be considered as a small 
fluid volume (or Lagrangian discretization point) characterized by a position $\bm{r}_i$, velocity $\bm{v}_i$, and mass $m_i$. 
In addition, each SDPD particle possesses a spin angular velocity $\bm{\psi}_i$ and moment of inertia $I_i$ introduced for 
the enforcement of angular momentum conservation \cite{Mueller_SDPD_2015}. 

SDPD particles $i$ and $j$ interact through four pairwise forces, including conservative $\bm{F}_{ij}^C$,
dissipative $\bm{F}_{ij}^D$, rotational $\bm{F}_{ij}^R$, and random $\tilde{\bm{F}}_{ij}$ forces given by
\begin{equation}
 \begin{gathered}
   \bm{F}_{ij}^C = \left(\frac{\bm{\Pi}_i}{d_i^2} + \frac{\bm{\Pi}_j}{d_j^2} \right)F_{ij}\cdot \bm{r}_{ij}, \\
   \bm{F}_{ij}^D = - \gamma_{ij}\left[\bm{v}_{ij} + (\bm{e}_{ij}\cdot\bm{v}_{ij})\bm{e}_{ij}\right], \\
   \bm{F}_{ij}^R = - \gamma_{ij}\frac{\bm{r}_{ij}}{2} \times (\bm{\psi}_i + \bm{\psi}_j), \\
   \tilde{\bm{F}}_{ij} =\sigma_{ij} \left( d\overline{\bm{\mathcal{W}}}^s_{ij} + \frac{1}{3} \mathrm{tr}[d \bm{\mathcal{W}}_{ij}] 
   \bm{1}\right) \cdot \frac{\bm{e}_{ij}}{dt},
 \end{gathered}
 \label{eq:forces}
\end{equation}
where $\bm{r}_{ij} = \bm{r}_i - \bm{r}_j$, $\bm{v}_{ij} = \bm{v}_i - \bm{v}_j$, and $\bm{e}_{ij} = \bm{r}_{ij}/r_{ij}$. 
Particle number density $d_i$ is computed as $d_i=\sum_j W_{ij}$ using a smoothing kernel function $W_{ij} = W(r_{ij})$ that vanishes beyond 
a cutoff radius $r_c$ and defines a non-negative function $F_{ij}$ through the equation $\nabla_i W_{ij} = -\bm{r}_{ij} F_{ij}$.
Then, particle mass density is given by $\rho_i = m_i d_0$. 
In the SDPD formulation for Newtonian fluids, the stress tensor $\bm{\Pi}_i = p_i^\mathrm{s}\bm{1}$ contains only diagonal (i.e. pressure)
components. The pressure $p_i^\mathrm{s}$ is determined by the equation of state (EoS) $p_i^\mathrm{s} = p_0(d_i/d_0)^\nu-p_\mathrm{b}$, where $d_0$ is the 
average number density, and $p_0$, $\nu$, and $p_\mathrm{b}$ are freely selected parameters. Furthermore, $\gamma_{ij}$ and $\sigma_{ij}$ 
are the corresponding force amplitudes            
\begin{equation}
  \gamma_{ij} = \frac{20\eta}{7} \frac{F_{ij}}{d_i d_j}, \quad \sigma_{ij} = 2\sqrt{k_\mathrm{B}T\gamma_{ij}},
\end{equation}
where $\eta$ is the fluid dynamic viscosity and $T$ is the equilibrium temperature. Eq.~(\ref{eq:forces}) also contains 
a matrix of independent Wiener increments $d\bm{\mathcal{W}}_{ij}$ with its trace $\mathrm{tr}[d \bm{\mathcal{W}}_{ij}]$ and the 
traceless symmetric part $d\overline{\bm{\mathcal{W}}}^s_{ij} = \frac{1}{2}\left( d\bm{\mathcal{W}}_{ij} + d\bm{\mathcal{W}}_{ji}
\right) - \frac{1}{3} \mathrm{tr}[d \bm{\mathcal{W}}_{ij}]$, and the time step $dt$.      

The evolution of particle positions, translational and angular velocities is obtained by integration of the following equations 
of motion 
\begin{equation}
  \begin{gathered}
    \dot{\bm{r}}_i = \bm{v}_i,  \\
    m_i\dot{\bm{v}}_i = \sum_j \bm{F}_{ij} = \sum_j (\bm{F}_{ij}^C + \bm{F}_{ij}^D + \bm{F}_{ij}^R + \tilde{\bm{F}}_{ij}), \\ 
    \dot{\bm{\psi}}_i = \frac{1}{2I_i}\sum_j \bm{r}_{ij}\times \bm{F}_{ij}, 
  \end{gathered}
  \label{eq:sdpd}
\end{equation}
using the velocity-Verlet algorithm \cite{Allen_CSL_1991}. 

Fluid elasticity is introduced following the idea that every fluid particle contains $N_\mathrm{p}$ 
bead-spring dumbbells \cite{Quesada_VEF_2009}. Dumbbells are not explicitly modeled, but represented by a conformation tensor 
$\bm{c}$ that characterizes their stretching state within each particle. The conformation tensor is expressed as $\bm{c}_i =
1/N_\mathrm{p} \sum_a^{N_\mathrm{p}}\bm{q}_a\bm{q}_a$, where $\bm{q}_a$ is the end-to-end distance of the \textit{a}-th 
dumbbell within a fluid particle $i$. Then, the stress tensor $\bm{\Pi}_i$ in Eq.~(\ref{eq:forces}) is modified by the addition 
of $\bm{c}$ contribution as follows \cite{Quesada_VEF_2009} 
\begin{equation}
\bm{\Pi}_i = p_i^\mathrm{s}\bm{1}+N_\mathrm{p} d_i k_\mathrm{B}T (\bm{1}-\bm{c}_i). 
\end{equation}
Evolution of the conformation tensor proceeds according to \cite{Quesada_VEF_2009}
\begin{equation}
 \dot{\bm{c}}_i^{\mu \mu'} = \frac{1}{d_i}\bm{c}_i^{\mu \nu} \bm{\kappa}_i^{\nu\mu'} + \frac{1}{d_i}\bm{c}_i^{\mu'\nu} 
 \bm{\kappa}_i^{\nu\mu} + \frac{1}{\tau}(\delta^{\mu \mu'} - \bm{c}_i^{\mu\mu'}) + \frac{d\tilde{\bm{c}}_{i}^{\mu\mu'}}{dt},
 \end{equation}
where $\bm{\kappa}^{\mu\nu}_i = \sum_j F_{ij}\bm{r}_{ij}^\mu\bm{v}_{ij}^{\nu}$ is the velocity gradient tensor, $\tau$ is the dumbbell 
relaxation time, and $d\tilde{\bm{c}}$ is the noise term. This model corresponds to the well-known viscoelastic Oldroyd-B model, 
in which the total fluid viscosity $\eta=\eta_\mathrm{s}+\eta_\mathrm{p}$ has two contributions, including solvent $\eta_\mathrm{s}$ and polymer $\eta_\mathrm{p}$ components. 
The polymer contribution is given by $\eta_\mathrm{p}=k_\mathrm{B}T d_0 N_\mathrm{p}\tau$, and can easily be adjusted through the parameters  $N_\mathrm{p}$
and $\tau$. 

In this work, the smoothing kernel is represented by the two dimensional (2D) Lucy function \cite{Lucy_NAT_1977}
\begin{equation}
W(r) = \frac{5}{\pi r_c^2} \left( 1 + 3\frac{r}{r_c}\right) \left( 1 - \frac{r}{r_c}\right)^3. 
\end{equation}
Thermal fluctuations are neglected by setting $k_\mathrm{B}T=10^{-6}$, such that the SDPD method is essentially reduced to SPH. 
Furthermore, we also neglect the noise term of the conformation tensor, i.e. $d\tilde{\bm{c}}=0$.

\subsection{Sheet model and simulation setup}

Figure \ref{fig:01}(a) shows a schematic of our 2D simulation setup with two sheets. According to the theoretical work of Taylor
\cite{Taylor_ASO_1951}, traveling wave $y(x,t) = b\sin(kx-\omega t + \phi)$ on an inextensible 2D sheet can be modeled through 
the imposition of particle velocities as
\begin{equation}
  \begin{gathered}
    v_x = \frac{\omega}{k}-Q\cos\theta, \quad  v_y = -Q\sin\theta, \\
    \tan\theta = bk\cos(kx-\omega t + \phi), \quad
    Q = \frac{\omega}{2\pi k}\int_0^{2\pi} \left(1+b^2k^2\cos^2\xi\right)^{1/2} \mathrm{d}\xi,
  \end{gathered}
  \label{eq:inextensible}
\end{equation}
where $b$ is the wave amplitude, $k$ is the wave number related to the wave length $\lambda=2\pi/k$, $\omega$ is the wave angular frequency,
and $\phi$ is the phase shift. Even though this traveling wave propagates with a wave speed $\omega/k$, material points of the sheet do 
not move forward or backward on average, and thus they represent a waving (rather than swimming) sheet, which will be referred to 
as {\it prescribed actuation} further in text. However, such sheet actuation generates a far-field flow \cite{Taylor_ASO_1951}, 
which can result in non-zero hydrodynamic synchronization forces between two waving sheets.

The prescribed actuation strategy in Eq.~(\ref{eq:inextensible}) cannot be used to model dynamic synchronization of two swimming sheets.
Furthermore, it does not account for a possible flexural rigidity of the sheets. Model of a flexible sheet is shown in Fig.~\ref{fig:01}(b),
where three layers of sheet particles are interconnected by harmonic springs. The spring potential is given by 
\begin{equation}
U(l) = \frac{\zeta_\mathrm{s}}{2} \left( l - l_0 \right)^2,
\end{equation} 
where $\zeta_\mathrm{s}$ is the spring stiffness, $l$ is its length, and $l_0$ is the equilibrium spring length. Actuation of the flexible 
sheets is performed using the middle layer (marked red in Fig.~\ref{fig:01}(b)), where a harmonic angle potential 
\begin{equation}
U(\theta) = \frac{\zeta_\theta}{2} \left( \theta - \theta_0 \right)^2
\label{eq:angle}
\end{equation}
is implemented for each pair of adjacent springs. Here, $\zeta_\theta$ is the potential strength, $\theta$ is the instantaneous angle between two adjacent springs in 
the middle layer, and $\theta_0$ is the spontaneous angle. A traveling wave on the sheets is imposed by the varying 
$\theta_0(x,t) = \theta_b\sin(kx_s-\omega t)$, where $\theta_b$ is the angle amplitude and $x_s = i l_0$ is the distance along 
the sheet with $i$ representing particle numbering along the middle layer.  Flexural rigidity $\kappa$ of this model
can be estimated as $\kappa = 2 \zeta_\mathrm{s} l_0^3 + \zeta_\theta l_0$ (see Appendix). In addition to the parameters 
$\zeta_\theta$ and $\theta_b$, the actual wave amplitude in this case is affected by the sheet flexural rigidity, wave frequency, 
and fluid viscoelasticity.  This model of sheet motion will be referred to as {\it internal actuation}, and allows the 
simulation of dynamic synchronization of two swimming sheets.  To constrain sheet motion in the $y$ direction, 
a tether force $\mathbf{F}_\mathrm{teth} = -\zeta_\mathrm{teth}(y_\mathrm{cm}-y_0)\mathbf{e}_y$ is applied uniformly to 
all sheet particles, where $\zeta_\mathrm{teth}$ is the spring stiffness, $y_\mathrm{cm}$ is the center-of-mass position of the sheet, 
and $y_0$ is the preferred position along the $y$ axis. Since simulation domain is periodic, any rotation of the sheet would necessarily result 
in its stretching, and therefore, its average direction along the $x$ axis is self constrained.    

In simulations, two sheets separated by the distance $h$ between their average positions are embedded into the modeled SDPD fluid. 
Each sheet is constructed by three layers of particles and is driven either through the prescribed actuation [see Eq.~(\ref{eq:inextensible})]
or by internal actuation [see Eq.~(\ref{eq:angle})]. The number density of 
sheet particles is the same as that of fluid particles. Therefore, interactions between fluid and sheet particles are identical to 
the fluid-fluid interactions. The simulation domain $L_x \times L_y$ is periodic in both dimensions. 
The cutoff radius is fixed at $r_c=1.6$ in all simulations and will be used as a basic length scale. Also, fluid resolution 
characterized by the number density $d_0 = 16/ r_c^2$ is kept the same in all simulations. Furthermore, we introduce reference 
mass density $\rho_\mathrm{ref} = 6.25$ and dynamic viscosity $\eta_\mathrm{ref}=6.25$, which define a mass scale 
$m_\mathrm{ref} = \rho_\mathrm{ref} r_c^2$ and a time scale $t_\mathrm{ref}=r_c^2 \rho_\mathrm{ref} / \eta_\mathrm{ref}$.  
We also define $F_\mathrm{ref}=r_c^3 \rho_\mathrm{ref} /t_\mathrm{ref}^2$, $p_\mathrm{ref}=r_c^2 \rho_\mathrm{ref}/t_\mathrm{ref}^2$, 
$\zeta_\mathrm{ref} = r_c^2\rho_\mathrm{ref}/t_\mathrm{ref}^2$, $\zeta_\mathrm{ref}^\theta =r_c^4 \rho_\mathrm{ref}/t_\mathrm{ref}^2$, $\kappa_\mathrm{ref}=r_c^5 \rho_\mathrm{ref} /t_\mathrm{ref}^2$ for scaling force, 
pressure, spring stiffness, angle potential strength, and flexural rigidity units, respectively. Simulation parameters are summarized in Table~\ref{tab:par} or otherwise specified 
explicitly in text. 

\begin{table}
\centering
\caption{Parameters used in simulations.}
\label{tab:par}
\begin{tabular}{ l l }
\hline	
\textbf{Parameters} & \textbf{Values} \\
\hline
Cutoff radius $r_c$ & 1.6 \\
Reference mass density $\rho_\mathrm{ref}$ & 6.25 \\
Reference dynamic viscosity $\eta_\mathrm{ref}$ & 6.25 \\
Energy unit $k_BT$ & 1e-6 \\
Size of the simulation domain $L_x \times L_y$ & $12.5 r_c \times 18.75r_c$ \\
Wave number $k$ & $ 4 \pi / L_x$\\
Number density $d_0 $ & $16/r_c^2$ \\
Time unit $t_\mathrm{ref}$ & $r_c^2 \rho_\mathrm{ref} / \eta_\mathrm{ref}$ \\
Force unit $F_\mathrm{ref}$ & $r_c^3 \rho_\mathrm{ref} /t_\mathrm{ref}^2$ \\
Pressure unit $p_\mathrm{ref}$ & $r_c^2 \rho_\mathrm{ref}/t_\mathrm{ref}^2$ \\
Spring stiffness unit $\zeta_\mathrm{ref}$ & $r_c^2 \rho_\mathrm{ref}/t_\mathrm{ref}^2$ \\
Angle potential strength unit $\zeta^\theta_\mathrm{ref}$ & $r_c^4 \rho_\mathrm{ref}/t_\mathrm{ref}^2$ \\
Flexural rigidity unit $\kappa_\mathrm{ref}$& $r_c^5 \rho_\mathrm{ref} /t_\mathrm{ref}^2$ \\
Average distance between the two sheets $h$ & $2.875r_c$ \\
Stiffness of the tether spring $\zeta_\mathrm{teth}$ & $8192  \zeta_\mathrm{ref}$ \\
Angle potential strength $\zeta_\theta$ & $640 \zeta^\theta_\mathrm{ref}$ \\
Difference in pressure coefficients $p_0 - p_\mathrm{b}$ & $32.8p_\mathrm{ref}$ \\
\hline
\end{tabular}
\end{table}

\begin{figure}[!htb]
     \centering
     \includegraphics[width=0.9\linewidth]{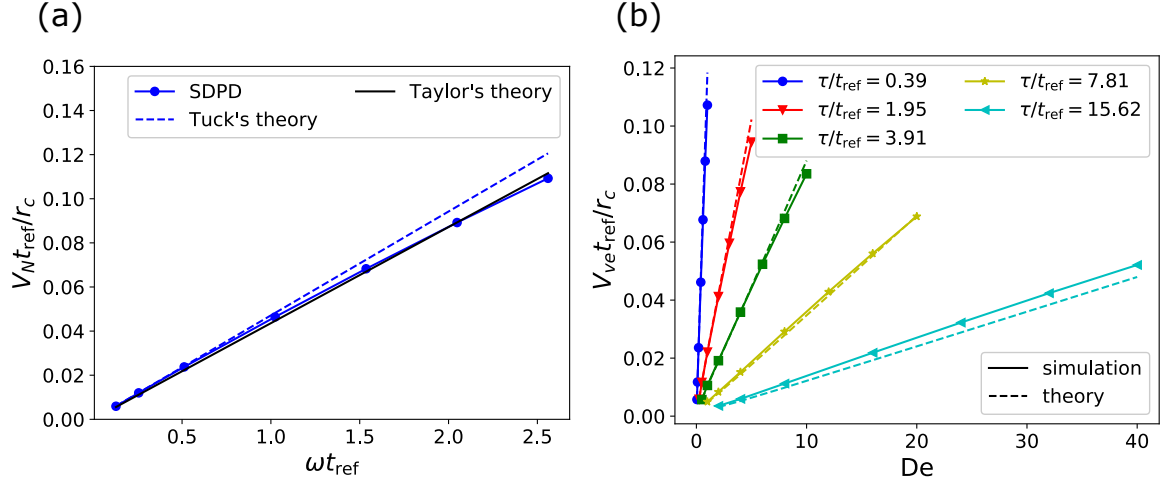}
     \caption{Model validation. (a) Far-field velocity $V_\mathrm{N}$ generated by a single waving sheet in a Newtonian fluid in comparison with the theoretical 
     	prediction by Taylor \cite{Taylor_ASO_1951} for $\mathrm{Re}=0$ [Eq.~(\ref{eq:taylor})] and  by Tuck \cite{Tuck_NSP_1968} for $\mathrm{Re}>0$. 
     	Here, $\eta/\eta_\mathrm{ref}=64$. (b) Comparison of simulated and 
     theoretical far-field velocities $V_\mathrm{ve}$ generated by a waving sheet in viscoelastic fluids with various relaxation times $\tau$. The theoretical 
     predictions correspond to Eq.~(\ref{eq:lauga}). Here, $N_\mathrm{p}=3\times 10^6$ and $\eta_\mathrm{s}/\eta_\mathrm{ref}=64$. In all simulations, 
     $L_x\times L_y = 18.75r_c\times 31.25r_c$, $b/r_c=0.375$, $\rho/\rho_\mathrm{ref}=1$, $p_0/p_\mathrm{ref}=409.6$, and $\nu=7$.}
     \label{fig:02}
\end{figure}

\subsection{Model validation}

Since our SDPD formulation is practically reduced to SPH by neglecting random terms, we can take advantage of rich SPH literature for the method validation. 
For instance, a similar SPH formulation has been used to simulate different fluid flows at low and moderate $\mathrm{Re}$ \cite{Morris_MLR_1997,Sigalotti_SPH_2003},
in good agreement with the corresponding analytical and/or finite-element results. Note that SPH may become unstable at high $\mathrm{Re}$ \cite{Meister_RNS_2014}. 
To verify the correctness of SDPD implementation for Newtonian fluids, wall-bounded Poiseuille flow and unsteady flow above an oscillating plate (or Stokes second problem) 
were simulated, showing an excellent agreement with the corresponding analytical solutions. The viscoelastic Oldroyd-B model has also been validated using unsteady Kolmogorov flow 
\cite{Quesada_VEF_2009}. 

In the context of swimming sheets, we revisit the problem of a single waving sheet both in Newtonian and viscoelastic 
fluids. An inextensible 2D sheet actuated according to Eq.~(\ref{eq:inextensible}) generates a far-field flow 
velocity $V_\mathrm{N}$ in the $x$ direction. The theoretical prediction of $V_\mathrm{N}$ is given by \cite{Taylor_ASO_1951}
\begin{equation}
  V_\mathrm{N} = \frac{1}{2}\omega b^2k\left(1-\frac{19}{16}b^2k^2\right).
  \label{eq:taylor}
\end{equation}
Figure \ref{fig:02}(a) compares simulation results for $V_\mathrm{N}$ in a Newtonian fluid with the theoretical prediction 
in Eq.~(\ref{eq:taylor}), demonstrating an excellent agreement. Here, $\mathrm{Re} =  8.95\times 10^{-4}$, defined as 
$\mathrm{Re}= f b^2\rho/\eta$ with $f=\omega/(2\pi)$, is small enough and can be neglected. The theoretical result by Tuck \cite{Tuck_NSP_1968} 
for $\mathrm{Re}>0$ is also shown in Fig.~\ref{fig:02}(a); however, it is only of the order $\mathcal{O}(b^2k^2)$ and is therefore less accurate. 

For viscoelastic Oldroyd-B fluids, a theoretical prediction for the far-field velocity $V_\mathrm{ve}$ generated by a waving sheet is given by \cite{Lauga_PVF_2007} 
\begin{equation}
  V_\mathrm{ve} = \frac{1}{2}\omega b^2k \frac{1+\mathrm{De}^2\eta_\mathrm{s}/\eta}{1+\mathrm{De}^2},
  \label{eq:lauga}
\end{equation}
where $\eta_\mathrm{s}$ is the solvent component of viscosity and $\mathrm{De} = \tau \omega$ is the Deborah number that represents a ratio of 
relaxation time to the characteristic time of sheet motion. Figure \ref{fig:02}(b) shows the comparison of simulated $V_\mathrm{ve}$ for a waving sheet in 
various viscoelastic fluids against the analytical predictions in Eq.~(\ref{eq:lauga}) as a function of $\mathrm{De}$. The simulation results
(solid lines) for various $\tau$ agree well with the analytical predictions (dashed lines). A small deviation mainly results from a slight shear 
dependence of SDPD fluid viscosity. 

\begin{figure}[!htb]
	\centering
	\includegraphics[width=0.9\linewidth]{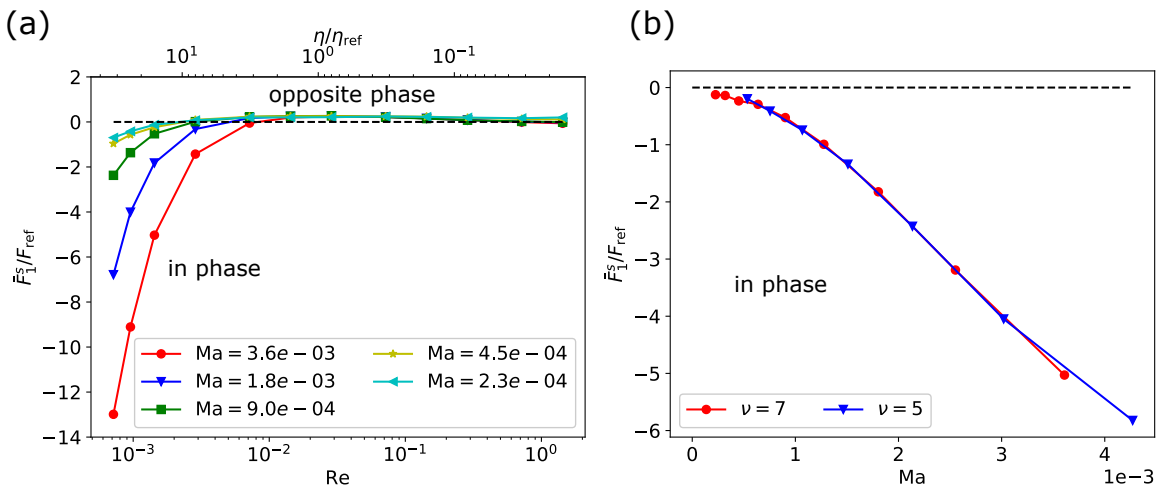}
	\caption{Synchronization force amplitudes of two waving sheets with prescribed actuation for different $\mathrm{Re}$ and $\mathrm{Ma}$ 
	numbers in a Newtonian fluid. (a) $\bar{F}^s_1$ as a 
	function of $\mathrm{Re}$ (fluid viscosity $\eta$ is varied) for various $\mathrm{Ma}$ values and $\nu = 7$. (b) $\bar{F}^s_1$ 
	as a function of $\mathrm{Ma}$ (the pressure parameter $p_0$ is varied) for $\nu = 5$ and $\nu = 7$. Here, $\eta/\eta_\mathrm{ref}=16$ and $\mathrm{Re}\approx 1.4\times 10^{-3}$.
	Note that $\bar{F}^s_1$ corresponds to the first sheet, while $\bar{F}^s_2=-\bar{F}^s_1$. In all cases, $\omega t_\mathrm{ref}=1.024$, $b/r_c=0.375$, and $\rho/\rho_\mathrm{ref}=1$.}
	\label{fig:03}
\end{figure}

\section{Results and discussion}

We investigate the synchronization of two sheets, and in particular, its dependence on inertial effects, fluid compressibility, sheet 
flexural rigidity, and fluid viscoelastic properties. Two sheets placed side by side with a distance $h$ apart (see Fig.~\ref{fig:01}) can 
have a phase difference in their motion given by $\phi_\mathrm{d} = \phi_2-\phi_1$. Generally, the synchronization force in the $x$ 
direction is a function of $\phi_\mathrm{d}$ and has a functional form \cite{Elfring_SFS_2011,Elfring_2DS_2010}
\begin{equation}
F^s(\phi_\mathrm{d}) = \bar{F}^s\sin(\phi_\mathrm{d}),
\label{eq:f_amp}
\end{equation} 
where $\bar{F}^s$ is the force amplitude. The synchronization forces on the two sheets have the same magnitude, but different 
signs, which means that they act in opposite directions. For the calculation of force amplitude $\bar{F}^s$, several simulations 
for different $\phi_\mathrm{d}$ values in the interval $[0,\pi]$ (with an increment of $\pi/30$ for simulations with Newtonian fluids and of 
$\pi/15$ for simulations with Oldroyd-B fluids) are performed, and the resultant force data 
are fitted using Eq.~(\ref{eq:f_amp}). Note that the force amplitude $\bar{F}^s$ can also be negative, as the fitting is carried out
within the range $[0,\pi]$. 

There exist two possible synchronized configurations: 
\begin{itemize}
\item [(i)] $\phi_\mathrm{d} = 0$ -- an in-phase configuration,
\item [(ii)] $\phi_\mathrm{d} = \pi$ -- an opposite-phase configuration.
\end{itemize}
Due to our definition of 
the phase difference as $\phi_\mathrm{d} = \phi_2-\phi_1$, the synchronization forces $F^s_1(\phi_\mathrm{d}) = -F^s_2(\phi_\mathrm{d}) < 0$
drive the sheets toward the in-phase configuration with $\phi_\mathrm{d} = 0$, while for $F^s_1(\phi_\mathrm{d}) = -F^s_2(\phi_\mathrm{d}) > 0$,
the opposite-phase conformation with $\phi_\mathrm{d} = \pi$ is stable. Further, simulation results will mainly be presented in terms of 
$\bar{F}^s_1$ for the first sheet only. 

\begin{figure}[!htb]
     \centering
     \includegraphics[width=0.9\linewidth]{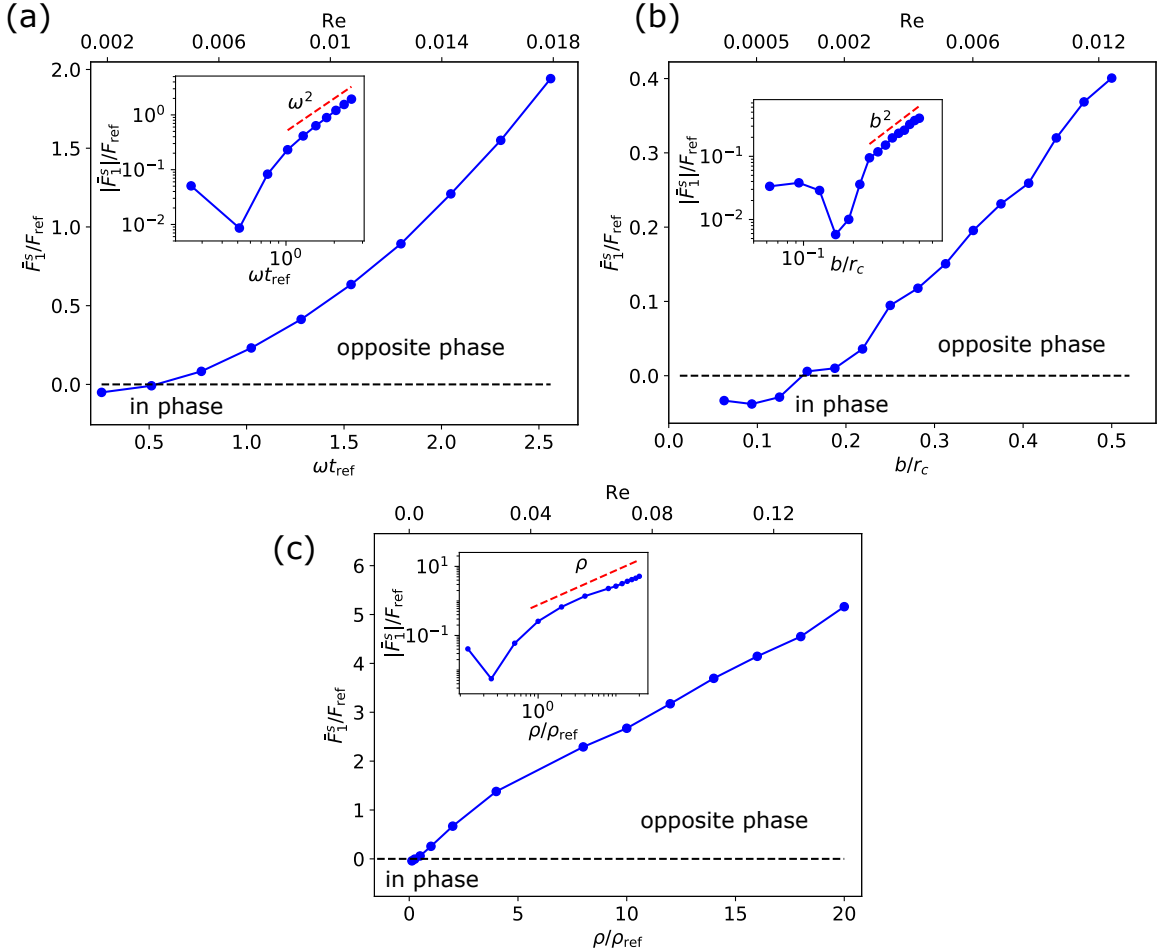}
     \caption{Synchronization force amplitudes of the first waving sheet (prescribed actuation) for various parameters. 
     (a) $\bar{F}^s_1$ as a function of wave frequency $\omega$. $b/r_c =0.375$ and $\rho/\rho_\mathrm{ref}=1$. (b) $\bar{F}^s_1$ as a function
     of wave amplitude $b$. $\omega t_\mathrm{ref}=1.024$ and $\rho/\rho_\mathrm{ref}=1$. (c) $\bar{F}^s_1$ as a function of fluid mass density $\rho$. 
     $\omega t_\mathrm{ref}=1.024$ and $b/r_c=0.375$. Other parameters, such as $\eta / \eta_\mathrm{ref}=3.2$, $p_0/ p_\mathrm{ref}=2621.4$, and $\nu=7$, are fixed in all simulations. 
     Insets show absolute values of $\bar{F}^s_1$ in log scale.}
     \label{fig:04}
\end{figure}

\subsection{Synchronization in Newtonian fluids}
\label{sec:newton}

\subsubsection{Interaction of two waving sheets} 

Figure \ref{fig:03}(a) presents synchronization force amplitudes of the first waving sheet with prescribed actuation for different values 
of $\mathrm{Re}$, which is controlled by changing the fluid viscosity $\eta$. For large enough $\mathrm{Re}$ (or small $\eta$), 
the synchronization forces favor the opposite-phase configuration, and appear to be independent of $\eta$. However, for small 
$\mathrm{Re}$ values, $\bar{F}^s_1$ becomes negative, favoring the in-phase configuration. Clearly, the magnitude of $\bar{F}^s_1$
at small $\mathrm{Re}$ depends on fluid compressibility, which is characterized by the Mach number $\mathrm{Ma} = f b/c$, 
where $c = (p_0 \nu/ \rho)^{1/2}$ is the speed of sound in SDPD fluid. The importance of fluid compressibility at large values 
of $\eta$ can be understood through a slow fluid relaxation by viscous diffusion in response to the prescribed 
sheet actuation that is independent of $\eta$. Figure \ref{fig:03}(b) shows that the magnitude of $\bar{F}^s_1$ at large $\eta$ 
is significantly reduced with decreasing $\mathrm{Ma}$, because local fluid relaxation by sound wave propagation becomes faster
for less compressible fluids. 

When $\mathrm{Ma}$ is decreased, the transition from opposite-phase to in-phase stable configuration in Fig.~\ref{fig:03}(a) is 
shifted toward lower $\mathrm{Re}$ values, and the magnitude of $\bar{F}^s_1$ in Fig.~\ref{fig:03}(b) is significantly reduced.
Furthermore, in the limit of incompressible fluid at $\mathrm{Re}=0$, no synchronization (i.e. $\bar{F}^s_1=0$) should occur 
for two interacting sheets having a reflection symmetry with respect to the $y$ axis \cite{Elfring_HPL_2009}, which is the case 
for the imposed sine wave here. These arguments suggest that for an incompressible fluid, inertial effects (i.e., $\mathrm{Re}>0$) 
should always lead to the stable opposite-phase configuration with $\bar{F}^s_1 > 0$ that vanishes at $\mathrm{Re}=0$. 
However, if fluid compressibility becomes relevant, the in-phase 
configuration may occur at low $\mathrm{Re}$. For comparison, human sperm typically has a beating 
frequency of $f\approx 20\,\mathrm{Hz}$ and an amplitude of $b \approx 10\,\mathrm{\mu m}$  \cite{Gaffney_ARFM_2011,Saggiorato_HSS_2017}, 
resulting in $\mathrm{Re} \approx 2\times 10^{-3}$ and $\mathrm{Ma} \approx 10^{-7}$ in a water-like environment. 
For existing microrobots, typical non-dimensional numbers are $\mathrm{Re} \sim 1$ and $\mathrm{Ma} \sim 10^{-5} - 10^{-4}$ \cite{Diller_CDM_2014,Huang_SMM_2016}. 
Thus, inertial effects are expected to be pertinent for all microswimmers, while fluid compressibility can be neglected for biological 
microswimmers such as sperm, but may become relevant for microrobots.     

\begin{figure}[!htb]
  \centering
  \includegraphics[width=0.45\linewidth]{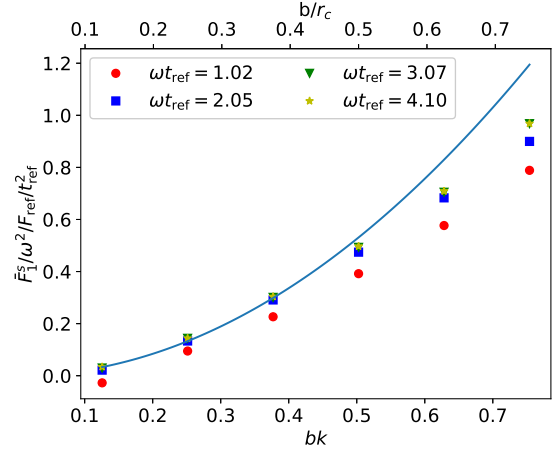}
  \caption{Synchronization force amplitudes with non-negligible inertia as a function of $bk$. The solid line represents a fit with 
  quadratic function. Note that for $bk>0.4$, the relation $\bar{F}^s_1 \propto b^2$ no longer holds. Here, $\eta / \eta_\mathrm{ref}
  =3.2$, $\rho/\rho_\mathrm{ref} =1$, $p_0/p_\mathrm{ref}=2621.4$, and $\nu=7$.} 
  \label{fig:05}
\end{figure}

To systematically investigate the synchronization of inextensible sheets, different parameters (other than $\eta$) that affect $\mathrm{Re}$ 
are varied. Figure~\ref{fig:04} shows synchronization force amplitudes of the first waving sheet (prescribed actuation) as a 
function of wave frequency $\omega$, wave amplitude $b$, and fluid mass density $\rho$. As discussed above, 
the existence of stable in-phase configuration for all varied parameters is due to fluid-compressibility effects at low $\mathrm{Re}$. 
When inertial effects dominate at large enough $\mathrm{Re}$, the opposite-phase conformation takes place, and the magnitude 
of $\bar{F}^s_1$ increases significantly with increasing $\mathrm{Re}$. 
Insets in Fig.~\ref{fig:04} show that at large enough $\mathrm{Re}$, the synchronization force amplitude $\bar{F}^s_1$ 
exhibits a power-law dependence with respect to the wave frequency $\omega$, wave amplitude $b$, and fluid mass density 
$\rho$. Remember that $\bar{F}^s_1$ is independent of fluid viscosity $\eta$, when fluid compressibility effects can be 
neglected, see Fig.~\ref{fig:03}(a). Therefore, we hypothesize that 
\begin{equation}
\bar{F}^s_1 \propto \rho \omega^2 b^2
\label{eq:synf}
\end{equation} 
for large enough $\mathrm{Re}$. Note that Eq.~(\ref{eq:synf}) is valid only for small $bk$. Figure \ref{fig:05} presents 
$\bar{F}^s_1$ as a function of $bk$ for various $\omega$, and shows that for $bk>0.4$, the increase in $\bar{F}^s_1$ becomes 
slower than $b^2$. As we will show later, an increase in the synchronization force amplitude for large $bk$ can be much 
faster than $b^2$ for viscoelastic fluids.  

\begin{figure}[!htb]
     \centering
     \includegraphics[width=0.9\linewidth]{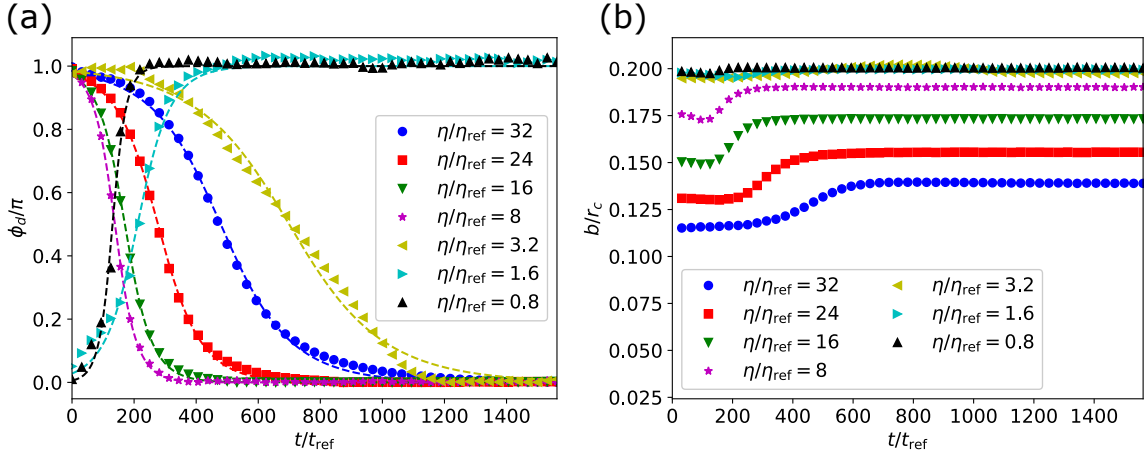}
     \caption{Dynamic synchronization of two flexible sheets (internal actuation) for different fluid viscosities. 
     (a) Phase difference plotted by symbols as a function of time. The dashed lines correspond to data fitting using 
     Eq.~(\ref{eq:syn_s}). (b) Beating wave amplitude $b$ for various $\eta$. Here, $\omega t_\mathrm{ref}=2.56$, $\theta_b/ \pi=0.028$, 
     $\zeta_\mathrm{s}/ \zeta_\mathrm{ref}=4096$ ($\kappa/\kappa_\mathrm{ref}=288$), $\rho/\rho_\mathrm{ref}=1$, $p_0/p_\mathrm{ref}=40.96$, and $\nu=5$.}
     \label{fig:06}
\end{figure}

\subsubsection{Dynamic synchronization of two flexible sheets}

To investigate dynamic synchronization process of two beating sheets, we employ the setup with two flexible sheets that 
have internal actuation. Figure \ref{fig:06}(a) shows time-dependent phase difference $\phi_\mathrm{d}$ between two flexible sheets
for various fluid viscosities. At low $\eta$, the sheets attain the opposite-phase configuration because of inertial effects at 
large enough $\mathrm{Re}$. As $\eta$ is increased, the in-phase configuration becomes stable due to the combined effect of fluid 
compressibility mentioned above and sheet flexibility that will be discussed below. Note that in the case of internal actuation the wave 
amplitude $b$ of flexible sheets is reduced significantly by increasing $\eta$ [see Fig.~\ref{fig:06}(b)] because of an increased viscous 
resistance on the sheets. An increase in viscosity and reduction in the wave amplitude are expected to decrease synchronization forces and 
increase synchronization time $\tau^\mathrm{s}$ \cite{Elfring_PHS_2011}. Nevertheless, the behavior of $\tau^\mathrm{s}$ in Fig.~\ref{fig:06}(a)
is non-monotonic with $\eta$ due to several reasons. In case of stable opposite-phase configuration ($\eta/ \eta_\mathrm{ref} \lesssim 2$ here), 
where fluid inertial effects dominate, $\tau^\mathrm{s}$ increases with increasing $\eta$ because the synchronization force amplitude 
$\bar{F}^s$ is independent of $\eta$ [see Fig.~\ref{fig:03}(a)]. The case of $\eta/ \eta_\mathrm{ref}=3.2$ exhibits the largest $\tau^\mathrm{s}$, 
since it is near the opposite-phase to in-phase transition, at which the synchronization force vanishes. With a further increase of viscosity 
($\eta/ \eta_\mathrm{ref} > 5$) for the in-phase configuration, although the magnitude of $\bar{F}^s$ increases, frictional forces on the sheets 
lead to larger $\tau^\mathrm{s}$ values for larger $\eta$. 

\begin{figure}[!htb]
	\centering
	\includegraphics[width=0.9\linewidth]{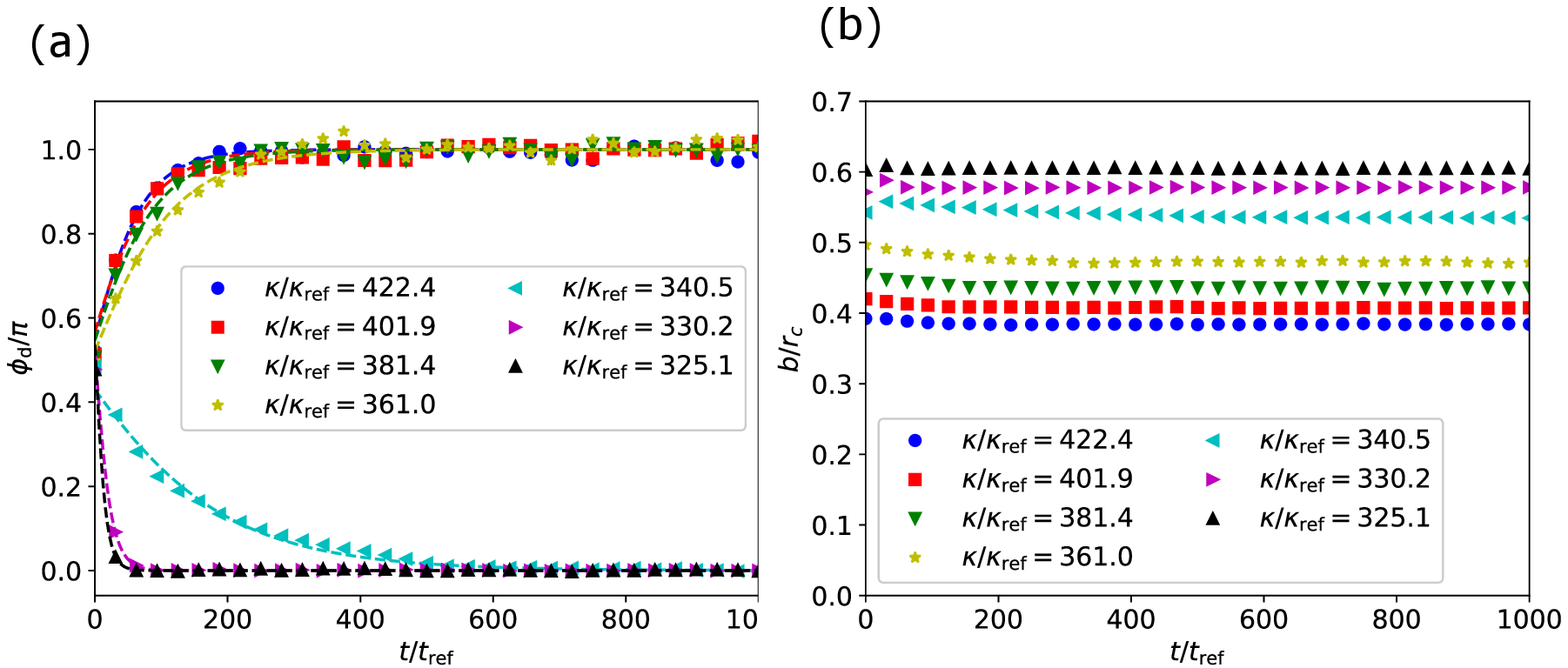}
	\caption{Synchronization of two flexible sheets for various flexural rigidities $\kappa$. 
		(a) Phase difference $\phi_\mathrm{d}$. Simulated data are shown by symbols, while the dashed lines represent fits using 
		Eq.~(\ref{eq:syn_s}). (b) Wave amplitude $b$. Here, $\zeta_\mathrm{s}/\zeta_\mathrm{ref}\in [41,819.2]$, $\omega t_\mathrm{ref}=2.048$, 
		$\eta / \eta_\mathrm{ref}=3.2$, $\rho/\rho_\mathrm{ref}=1$, $\theta_b/ \pi=0.044$, $p_0/ p_\mathrm{ref} =2621.4$, and $\nu=7$.}
	\label{fig:07}
\end{figure}

To directly demonstrate the effect of  sheet flexibility, Fig.~\ref{fig:07}(a) presents the transition from opposite-phase to in-phase configuration 
when sheet flexibility is increased by decreasing the spring stiffness $\zeta_\mathrm{s}$. Here, fluid compressibility effects can be 
neglected due to a low fluid viscosity of $\eta / \eta_\mathrm{ref}=3.2$. Interestingly, stiff sheets (i.e., large $\zeta_\mathrm{s}$ and $\kappa$) 
synchronize toward the opposite-phase configuration due to inertial effects, while soft sheets are driven toward the in-phase conformation, 
even though the effective $\mathrm{Re}$ increases with decreasing $\kappa$ due to an increase in wave amplitude, see Fig.~\ref{fig:07}(b). 
Therefore, sheet flexibility drives the system of two sheets toward the in-phase configuration. This result is consistent with the theory predicting 
that a finite elasticity of two sheets in an incompressible fluid is sufficient to break the symmetry and exhibit the in-phase synchronization 
at $\mathrm{Re}=0$ \cite{Elfring_SFS_2011}. Furthermore, a simulation study \cite{Fauci_IOF_1990} of a pair of flexible sheets based 
on the incompressible Navier-Stokes equations reports the transition from in-phase to opposite-phase conformation with increasing $\mathrm{Re}$, 
in agreement with the discussed results. 

Our simulations demonstrate that the synchronization forces $F^s_1$ and $F^s_2$ have a sine-function dependence on 
$\phi_\mathrm{d}$ [see Eq.~(\ref{eq:f_amp})], independently of the contributing factors, such as fluid compressibility, inertial 
effects, and sheet flexibility. Therefore, the dynamic synchronization process can be described as a damped harmonic oscillator 
\begin{equation}
 \frac{d^2\phi_\mathrm{d}}{dt^2} = - a_1 \frac{d\phi_\mathrm{d}}{dt} - a_2\sin(\phi_\mathrm{d}),  
\label{eq:syn}
\end{equation}
where $a_1$ is a damping coefficient and $a_2$ corresponds to the synchronization force amplitudes $\bar{F}^s_1$ and $\bar{F}^s_2$, 
which are either negative or positive, depending on whether the in-phase or opposite-phase configuration is stable. In general, 
$a_1$ can be a function of $\phi_\mathrm{d}$, but for simplicity, it is assumed to be constant here. A process described by 
Eq.~(\ref{eq:syn}) can have an oscillating dynamics, if the inertial term represented by the second time derivative is large enough. 
Our simulations (not presented here) have shown that an oscillation in the synchronization of two flexible sheets may 
occur at low fluid viscosities $\eta / \eta_\mathrm{ref} < 0.08$. As the employed viscosity values $\eta / \eta_\mathrm{ref} \gg 0.08$,
the synchronization process of two sheets can be considered overdamped, even though it is often caused by inertial effects. 
By neglecting the inertial term in Eq.~(\ref{eq:syn}), an Adler-like equation \cite{Adler_SLP_1946} 
$d\phi_\mathrm{d} / dt = -a_2/a_1\sin(\phi_\mathrm{d})$ for $\phi_\mathrm{d}$ is obtained, which has an analytical solution given by
\cite{Niedermayer_SPL_2008}
\begin{equation}
  \phi_\mathrm{d} = 2\tan^{-1}\left(\tan\frac{\phi_\mathrm{d}^0}{2}e^{-a_2 t/a_1}\right),
\label{eq:syn_s}
\end{equation}
where $\phi_\mathrm{d}^0$ is the initial phase difference at time $t=0$. Equation~(\ref{eq:syn_s}) is used to fit the simulation data 
in Figs.~\ref{fig:06}(a) and \ref{fig:07}(a), where the fits are shown by dashed lines. Clearly, the fits are very good, and allow the extraction of 
synchronization time $\tau^\mathrm{s} = |a_1/a_2|$ from the simulation data of time-dependent synchronization. 

\begin{figure}[!htb]
   \centering
   \includegraphics[width=0.95\linewidth]{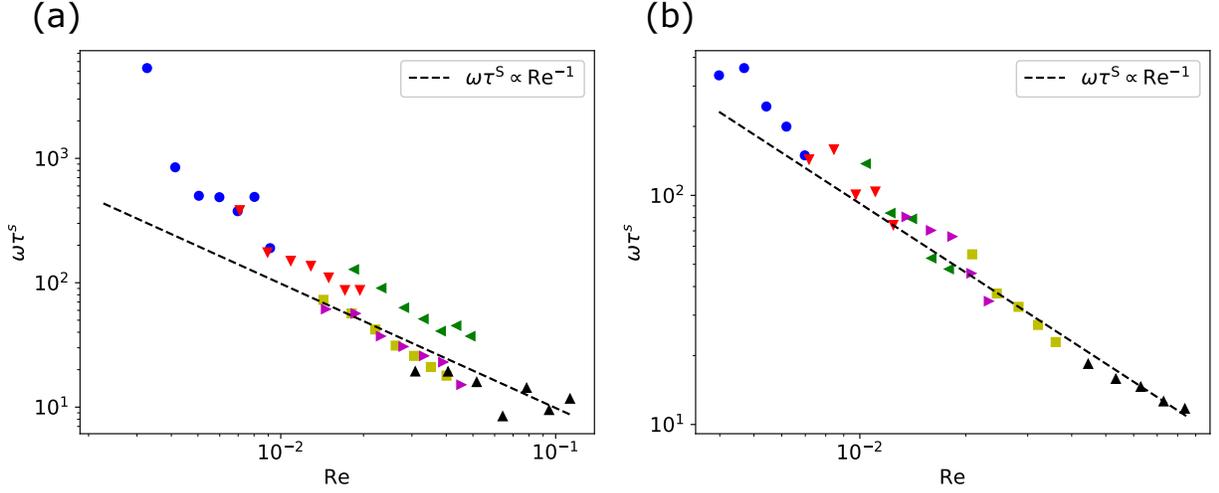}
   \caption{Dependence of synchronization time $\tau^\mathrm{s}$ on $\omega$ and $\mathrm{Re}$. (a) Simulations with fexural rigidity 
   	$\kappa/\kappa_\mathrm{ref}=221.2$ and the EoS parameter $p_0/ p_\mathrm{ref}=2621.4$. Different data sets are plotted by symbols, corresponding to 
   	$\eta / \eta_\mathrm{ref}=3.2$, $\rho/\rho_\mathrm{ref} = 1$, and $\theta_b / \pi = 0.028$ (blue circles);
   	$\eta / \eta_\mathrm{ref}=3.2$, $\rho/\rho_\mathrm{ref} = 1$, and $\theta_b / \pi = 0.044$ (red down-pointing triangles);
   	$\eta / \eta_\mathrm{ref}=3.2$, $\rho/\rho_\mathrm{ref} = 1$, and $\theta_b / \pi = 0.089$ (green left-pointing triangles);
   	$\eta / \eta_\mathrm{ref}=1.6$, $\rho/\rho_\mathrm{ref} = 1$, and $\theta_b / \pi = 0.044$ (yellow squares);
   	$\eta / \eta_\mathrm{ref}=3.2$, $\rho/\rho_\mathrm{ref} = 2$, and $\theta_b / \pi = 0.044$ (purple right-pointing triangles);
   	$\eta / \eta_\mathrm{ref}=3.2$, $\rho/\rho_\mathrm{ref} = 4$, and $\theta_b / \pi = 0.044$ (black up-pointing triangles).
   	(b) Simulations with $\kappa/\kappa_\mathrm{ref}=364.8$ and $p_0/ p_\mathrm{ref}=10485.6$. Various symbols represent 
   	$\eta / \eta_\mathrm{ref}=3.2$, $\rho/\rho_\mathrm{ref} = 1$, and $\theta_b / \pi = 0.044$ (blue circles);
   	$\eta / \eta_\mathrm{ref}=3.2$, $\rho/\rho_\mathrm{ref} = 1$, and $\theta_b / \pi = 0.067$ (red down-pointing triangles);
   	$\eta / \eta_\mathrm{ref}=3.2$, $\rho/\rho_\mathrm{ref} = 1$, and $\theta_b / \pi = 0.089$ (green left-pointing triangles);
   	$\eta / \eta_\mathrm{ref}=1.6$, $\rho/\rho_\mathrm{ref} = 1$, and $\theta_b / \pi = 0.089$ (yellow squares);
   	$\eta / \eta_\mathrm{ref}=3.2$, $\rho/\rho_\mathrm{ref} = 1$, and $\theta_b / \pi = 0.11$ (purple right-pointing triangles);
   	$\eta / \eta_\mathrm{ref}=3.2$, $\rho/\rho_\mathrm{ref} = 4$, and $\theta_b / \pi = 0.089$ (black up-pointing triangles).
   	Each set of data includes several $\omega t_\mathrm{ref}$ values in the range $[1.02,2.56]$ and $\nu=7$ in all simulations.
   	The dashed lines indicate $\mathrm{Re}^{-1}$.}
   \label{fig:08}
\end{figure}

Let us consider $a_1 \propto \eta$, since it represents damping effects, and $a_2=\bar{F}^s_1 \propto \rho \omega^2 b^2$ [see Eq.~(\ref{eq:synf})], 
which has been hypothesized for two waving sheets at high enough $\mathrm{Re}$. Then, the synchronization time $\tau^\mathrm{s}$ normalized
by the wave frequency $\omega$ simply becomes    
\begin{equation}
 \tau^\mathrm{s} \omega \propto \frac{1}{\mathrm{Re}}.
\label{eq:t_fre}
\end{equation}
This relation is tested by a number of simulations for various $\omega$, $\rho$, $\eta$, and $\theta_b$ values. Figure~\ref{fig:08}(a) 
shows $\tau^\mathrm{s}$ for the flexural rigidity $\kappa/\kappa_\mathrm{ref}=221.2$ and EoS parameter $p_0/ p_\mathrm{ref}=2621.4$ as a function 
of $\mathrm{Re}$. When inertial effects dominate at large enough $\mathrm{Re}$,  $\tau^\mathrm{s}$ is inversely proportional to 
$\mathrm{Re}$, as predicted. At low $\mathrm{Re}$ values, the effects of fluid compressibility and sheet flexibility become important, 
so that $\tau^\mathrm{s}$ deviates from this relation. Figure~\ref{fig:08}(b) presents simulation data for $\kappa/\kappa_\mathrm{ref}=364.8$ 
and $p_0/ p_\mathrm{ref}=10485.6$, such that both fluid compressibility and sheet flexibility effects are significantly reduced. The behavior 
of  $\tau^\mathrm{s}$ closely follows the relation in Eq.~(\ref{eq:t_fre}) for a wide range of $\mathrm{Re}$ numbers. 
Noteworthy, the rapid increase in $\tau^\mathrm{s}$ at small $\mathrm{Re}$ in Fig.~\ref{fig:08} is qualitatively consistent with the theoretical 
prediction that no synchronization of two inextensible sheets having a reflection symmetry with respect to the $y$ axis
can occur at $\mathrm{Re}=0$ \cite{Lauga_PVF_2007}. 

\begin{figure}[!htb]
  \centering
  \includegraphics[width=0.9\linewidth]{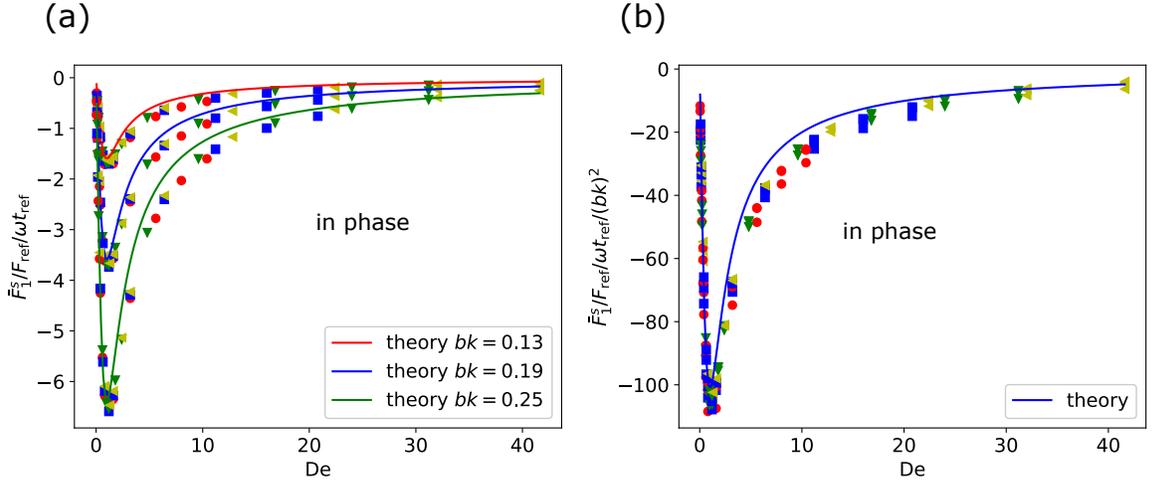}
  \caption{Synchronization force amplitudes of two waving sheets (prescribed actuation) in viscoelastic fluids. (a) $\bar{F}^s_1 / \omega$
  and (b) $\bar{F}^s_1 / \omega / (bk)^2$ as a function of $\mathrm{De}$ for various $bk$. Different simulation data sets are plotted by 
  symbols, corresponding to $\omega t_\mathrm{ref} =1.024$ (red circles); $\omega t_\mathrm{ref} =2.048$ (blue squares); 
  $\omega t_\mathrm{ref} =3.072$ (green down-pointing triangles); $\omega t_\mathrm{ref} =4.096$ (yellow left-pointing triangles).
  Solid lines are theoretical predictions \cite{Elfring_2DS_2010} from Eq.~(\ref{eq:el}).  
  Here, $\eta_\mathrm{s} / \eta_\mathrm{ref}  = 3.0$, $\eta_\mathrm{p} / \eta_\mathrm{ref} = 13.0$, and $\rho/\rho_\mathrm{ref} = 1$.} 
\label{fig:09}
\end{figure}

\subsection{Synchronization in viscoelastic fluids}
\label{sec:nonnewton}

Fluid elasticity is also sufficient to break the symmetry and result in the in-phase synchronization of two inextensible sheets at
$\mathrm{Re}=0$ \cite{Elfring_2DS_2010}. Theoretical prediction for the synchronization force between two sheets up to the order 
$\mathcal{O}(b^2k^2)$ is given by \cite{Elfring_2DS_2010}
\begin{equation}
 \frac{F^s_1(\phi_\mathrm{d}) k}{\omega \eta} = \left[\frac{2\pi \Delta U}{kh} - 
    \frac{4\pi \mathrm{De}\eta_\mathrm{p}}{\eta (1+\mathrm{De}^2)} A(kh) \sin(\phi_\mathrm{d})\right] (bk)^2, \quad
    A(kh) = \frac{kh\cosh (kh) + \sinh (kh)}{\cosh(2kh) - 2k^2h^2 -1},
 \label{eq:el}
\end{equation}
where $\Delta U$ is the relative velocity of two sheets and $\eta = \eta_\mathrm{p} + \eta_\mathrm{s}$ includes polymer and solvent 
viscosity contributions. In case of two waving sheets with prescribed actuation, $\Delta U = 0$. Note that for $\Delta U = 0$, 
the synchronization force in Eq.~(\ref{eq:el}) can be written as $F^s_1(\phi_\mathrm{d}) = \bar{F}_1^s\sin(\phi_\mathrm{d})$, which is 
identical to Eq.~(\ref{eq:f_amp}) hypothesized before.   

Figure \ref{fig:09}(a) presents the ratio $\bar{F}^s_1 / \omega$ obtained from a number of simulations (symbols) of two waving sheets
in viscoelastic fluids with respect to the theoretical prediction (solid lines) in Eq.~(\ref{eq:el}) for several $bk$ values. 
An excellent agreement between simulated and theoretical synchronization force amplitudes is achieved. In these simulations, the 
wave frequency $\omega t_\mathrm{ref} \in [1.024,4.096]$, wave amplitude $b/r_c \in [0.125,0.25]$, and the relaxation time 
$\tau/t_\mathrm{ref} \in [0.04,10.16]$ are varied, while the viscosities $\eta_\mathrm{s} / \eta_\mathrm{ref} = 3.0$ and 
$\eta_\mathrm{p} / \eta_\mathrm{ref} = 13.0$ are kept constant. Note that the value of $N_\mathrm{p}$ is adjusted according to changes in $\tau$ in order 
to keep $\eta_\mathrm{p} = k_\mathrm{B}T d_0 N_\mathrm{p} \tau$ fixed. Furthermore, mass density of the fluid is set to be relatively small 
$\rho/\rho_\mathrm{ref} = 0.0625$ to minimize inertial effects, and the EoS parameters ($p_0/ p_\mathrm{ref}=6553.6$ and $\nu=7$) are chosen such that 
fluid compressibility effects can be neglected. 

\begin{figure}[!htb]
	\centering
	\includegraphics[width=0.9\linewidth]{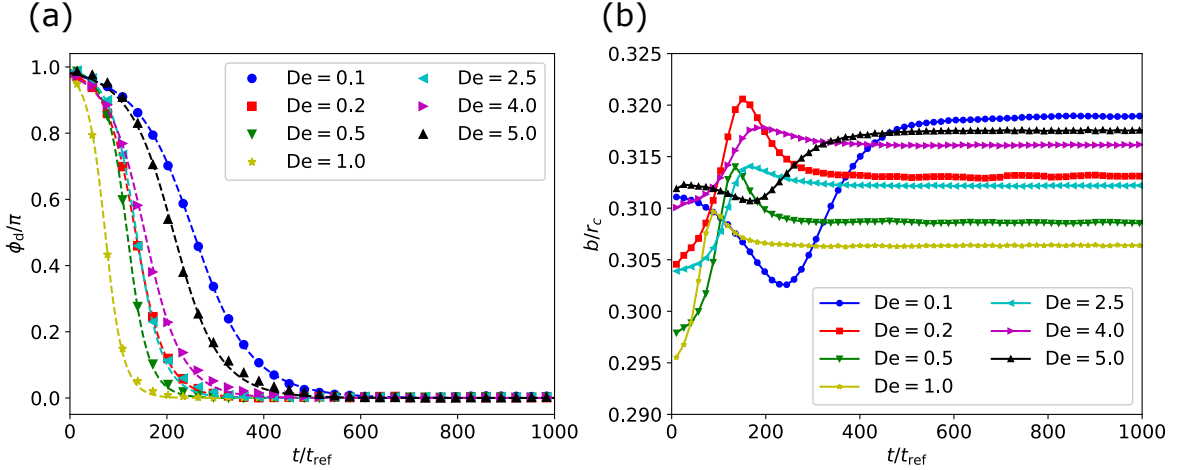}
	\caption{Dynamic synchronization of two flexible sheets mediated by fluid viscoelasticity. (a) Phase difference $\phi_\mathrm{d}$ as 
		a function of time for different $\mathrm{De}$. Dashed lines represent fits using Eq.~(\ref{eq:syn_s}). (b) Beating wave amplitudes. 
		Here, $\eta_\mathrm{s} / \eta_\mathrm{ref} = 3.0$, $\eta_\mathrm{p} / \eta_\mathrm{ref} = 13.0$, $\omega t_\mathrm{ref}=1.28$,
		$\rho/\rho_\mathrm{ref}=0.125$, $\theta_b / \pi=0.056$, $\zeta_\mathrm{s} / \zeta_\mathrm{ref}=4096$ ($\kappa/\kappa_\mathrm{ref}=288$), $p_0/ p_\mathrm{ref}=6553.6$, and $\nu=7$.} 
	\label{fig:10}
\end{figure}

Figure \ref{fig:09}(b) demonstrates that the simulated force amplitudes scaled as $\bar{F}^s_1 / \omega /(bk)^2$ fall onto a single master 
curve that is well captured by the theoretical prediction in Eq.~(\ref{eq:el}). Furthermore, Fig.~\ref{fig:09} shows that the maximum 
synchronization force is achieved at $\mathrm{De}=1$ for a fixed $bk$ and $\omega$. Noteworthy, for a fixed $\omega$, the synchronization 
force amplitude asymptotically approaches zero with increasing $\mathrm{De}$ or $\tau$. This indicates that the synchronization of 
flagellated microswimmers must be weak in viscoelastic fluids with a large relaxation time. In fact, $\mathrm{De}$ for realistic 
biological microswimmers can be significantly larger than unity. For instance, $\mathrm{De}$ is between $10^2$ and $10^3$ for the case 
of sperm cells swimming in mucus, whose relaxation time is in the range of $1\sim10\,\mathrm{s}$ \cite{Elfring_2DS_2010}.

Figure \ref{fig:10}(a) presents dynamic synchronization of two flexible sheets for various $\mathrm{De}$, and demonstrates that this 
process is fastest at $\mathrm{De}=1$, in agreement with the theoretical prediction in Eq.~(\ref{eq:el}). The corresponding beating wave 
amplitudes shown in Fig.~\ref{fig:10}(b) are small enough in these simulations, such that the theoretical prediction is accurate. 
Interestingly, the synchronization times in viscoelastic fluids have similar magnitudes as those in Newtonian fluids (compare with Figs.~\ref{fig:06} 
and \ref{fig:07}).  

\begin{figure}[!htb]
	\centering
	\includegraphics[width=0.5\linewidth]{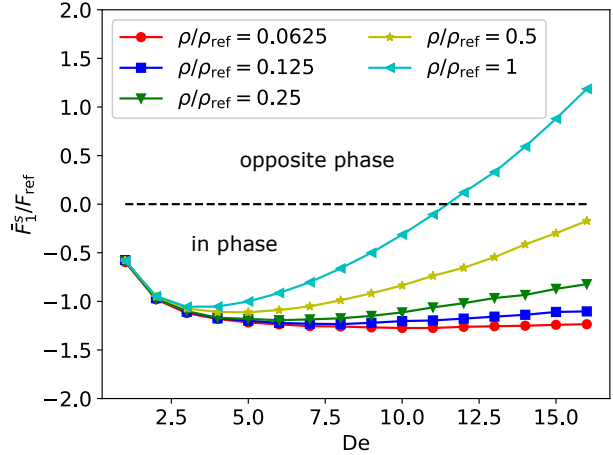}
	\caption{Synchronization force amplitudes of two waving sheets in viscoelastic fluids for different fluid densities, affecting 
		the value of $\mathrm{Re}$. $\mathrm{De}$ is varied by changing $\omega t_\mathrm{ref}\in[0.256,4.096]$. For instance, when 
		$\omega t_\mathrm{ref}=4.096$ ($\mathrm{De}=16$), $\mathrm{Re} = 1.6\times 10^{-4}$ for $\rho/\rho_\mathrm{ref} = 0.0625$ and 
		$\mathrm{Re} = 2.5\times 10^{-3}$ for $\rho/\rho_\mathrm{ref}=1$. Other parameters $b/r_c=0.25$, $\eta_\mathrm{s} / \eta_\mathrm{ref} = 6.0$, 
		$\eta_\mathrm{p} / \eta_\mathrm{ref} = 10$, $\tau / t_\mathrm{ref} = 3.9$, $p_0/ p_\mathrm{ref}=6553.6$, and $\nu=7$ remain fixed.} 
	\label{fig:11}
\end{figure}

According to the theoretical prediction in Eq.~(\ref{eq:el}) \cite{Elfring_2DS_2010}, the synchronization forces resulting from fluid 
viscoelasticity are of the order $\mathcal{O}(b^2k^2)$, which is similar to the synchronization forces originating from
inertial effects, see Eq.~(\ref{eq:synf}). Therefore, it is plausible to expect the transition from in-phase to opposite-phase 
configuration with increasing $\mathrm{Re}$ in viscoelastic fluids. Figure \ref{fig:11} illustrates the competing effects of 
viscoelasticity and inertia, and demonstrates the existence of in-phase to opposite-phase transition at high enough $\mathrm{Re}$ 
by increasing $\rho$. For negligible inertial effects ($\rho/\rho_\mathrm{ref}=0.0625$), $\bar{F}^s_1$ exhibits a plateau at large $\mathrm{De}$,
consistently with the theoretical prediction in Eq.~(\ref{eq:el}). As $\rho$ is increased, inertial effects become more prominent,
and the opposite-phase configuration might be attained. For comparison, when $\omega t_\mathrm{ref}=4.096$ ($\mathrm{De}=16$), $\rho/\rho_\mathrm{ref}=0.0625$ leads to 
$\mathrm{Re} = 1.6\times 10^{-4}$, while $\rho/\rho_\mathrm{ref}=1$ corresponds to $\mathrm{Re} = 2.5\times 10^{-3}$.  

\begin{figure}[!htb]
  \centering
  \includegraphics[width=0.9\linewidth]{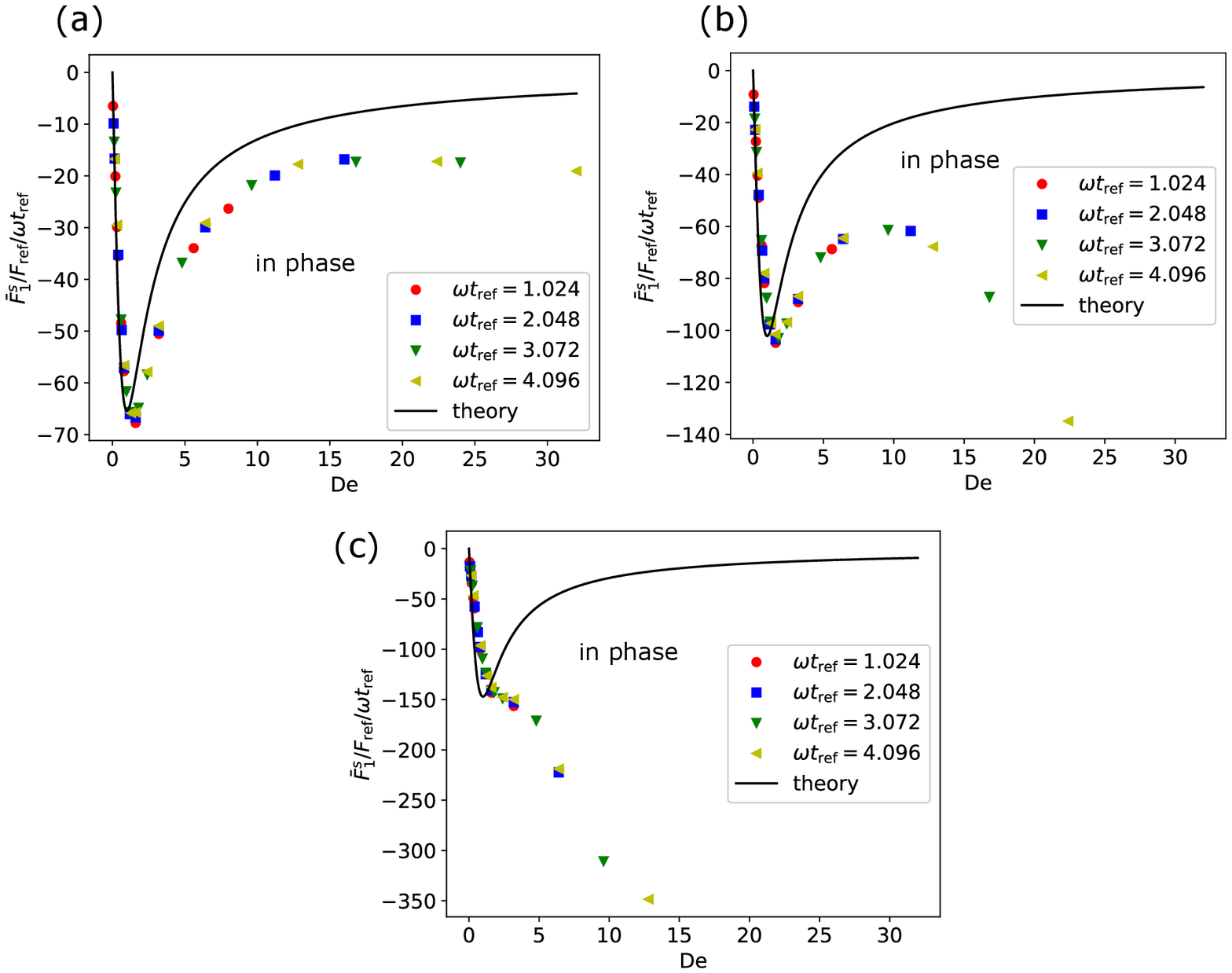}
  \caption{Synchronization force amplitudes as a function of $\mathrm{De}$ for (a) $bk = 0.5$, (b) $bk = 0.63$, and (c) $bk = 0.75$.
  As $bk$ is increased, synchronization forces strongly increase at large enough $\mathrm{De}$. Note that for large $bk$, 
  the dependence of $F^s$ on phase difference $\phi_\mathrm{d}$ is no longer a sine function. Therefore, $\bar{F}^s_1$ corresponds to
  the maximum of $F^s_1(\phi_\mathrm{d})$ here. Other parameters $\eta_\mathrm{s} / \eta_\mathrm{ref} = 3.0$, $\eta_\mathrm{p} / \eta_\mathrm{ref} 
  = 13.0$, $\rho/\rho_\mathrm{ref} = 0.0625$, $p_0/ p_\mathrm{ref}=6553.6$, and $\nu=7$ remain fixed.} 
  \label{fig:12}
\end{figure}

The prediction that synchronization of two sheets with a fixed $\omega$ is weak at large $\mathrm{De}$ seemingly disagrees with recent 
experimental \cite{Tung_FVE_2017} and numerical \cite{Ishimoto_HCS_2018} studies on sperm interaction, which demonstrate that 
fluid viscoelasticity can significantly enhance clustering. Note that the theoretical prediction by Taylor \cite{Taylor_ASO_1951} in 
Eq.~(\ref{eq:taylor}) is of the order $\mathcal{O}(b^4k^4)$, and has been shown to be accurate for $bk \lesssim 0.4$ 
\cite{Sauzade_TSS_2011}. The theoretical prediction in Eq.~(\ref{eq:el}) for Oldroyd-B fluids \cite{Elfring_2DS_2010} is of 
the order $\mathcal{O}(b^2k^2)$, such that it is reliable for small $bk$ as verified by our simulations in Fig.~\ref{fig:09}, but is expected to be 
less accurate for large $bk$ values. Furthermore, there exist a number of experimental studies
\cite{Woolley_SSF_2009,Nosrati_2DS_2015,Tung_FVE_2017,Saggiorato_HSS_2017} with flagellated
microswimmers, whose wave amplitude is large enough to make the assumption of small $bk$ invalid. Figure \ref{fig:12} shows synchronization 
force amplitudes of two waving sheets in viscoelastic fluids for $bk=0.5$, $bk=0.63$, and $bk=0.75$. The simulation parameters here are 
the same as those in Fig.~\ref{fig:09}. Clearly, the simulated $\bar{F}^s_1$ values for large $bk$ are significantly larger than those 
predicted theoretically at large $\mathrm{De}$. This demonstrates that for $bk > 0.4$, fluid viscoelasticity plays a much more prominent role for sheet 
synchronization than predicted theoretically for small $bk$. Interestingly, the dependence of $\bar{F}^s_1$ on $bk >0.4$ has an exponent that 
can be significantly larger than two. For comparison, an increase in the synchronization force for Newtonian fluids at large enough $\mathrm{Re}$ and 
$bk >0.4$ is slower than $b^2$, as shown in Fig.~\ref{fig:05}. Therefore, our simulations demonstrate that fluid viscoelasticity is the main cause of a tremendous 
increase in synchronization forces at large $bk$ and $\mathrm{De}$, providing a plausible explanation for the enhanced clustering of flagellated 
microswimmers in viscoelastic fluids. 

Note that the Oldroyd-B model becomes unphysical when the Weissenberg number, which relates fluid relaxation time to the time determined by a 
characteristic strain rate, approaches $\mathrm{Wi}\approx 1$. As a precautionary step, the Oldroyd-B model has been tested using Kolmogorov flow, 
leading to accurate results for $\mathrm{Wi} \lesssim 0.8$, while for $\mathrm{Wi} \gtrsim 0.9$ the SDPD simulations become unstable. 
For the synchronization problem of two sheets, Weissenberg number can be defined as $\mathrm{Wi} = bk \omega \tau/(2\pi)^2$. 
For example, simulation results in Fig.~\ref{fig:09} agree well with the analytical solution in Eq.~(\ref{eq:el}) and correspond to $\mathrm{Wi} \in [2.5\times 10^{-4},0.26]$
for $bk=0.25$. For comparison, $\mathrm{Wi} \in [5\times 10^{-4},0.41]$ for $bk=0.5$ in Fig.~\ref{fig:12}(a),  $\mathrm{Wi} \in [6.4\times 10^{-4},0.36]$ for $bk=0.63$ 
in Fig.~\ref{fig:12}(b), and $\mathrm{Wi} \in [7.6\times 10^{-4},0.24]$ for $bk=0.75$ in Fig.~\ref{fig:12}(c). Therefore, all presented results are well within the limit of 
Oldroyd-B model applicability, confirming that the dramatic increase in synchronization forces at large $\mathrm{De}$ and $bk$ is not due to any model shortcomings.  

\begin{figure}[!htb]
	\centering
	\includegraphics[width=0.8\linewidth]{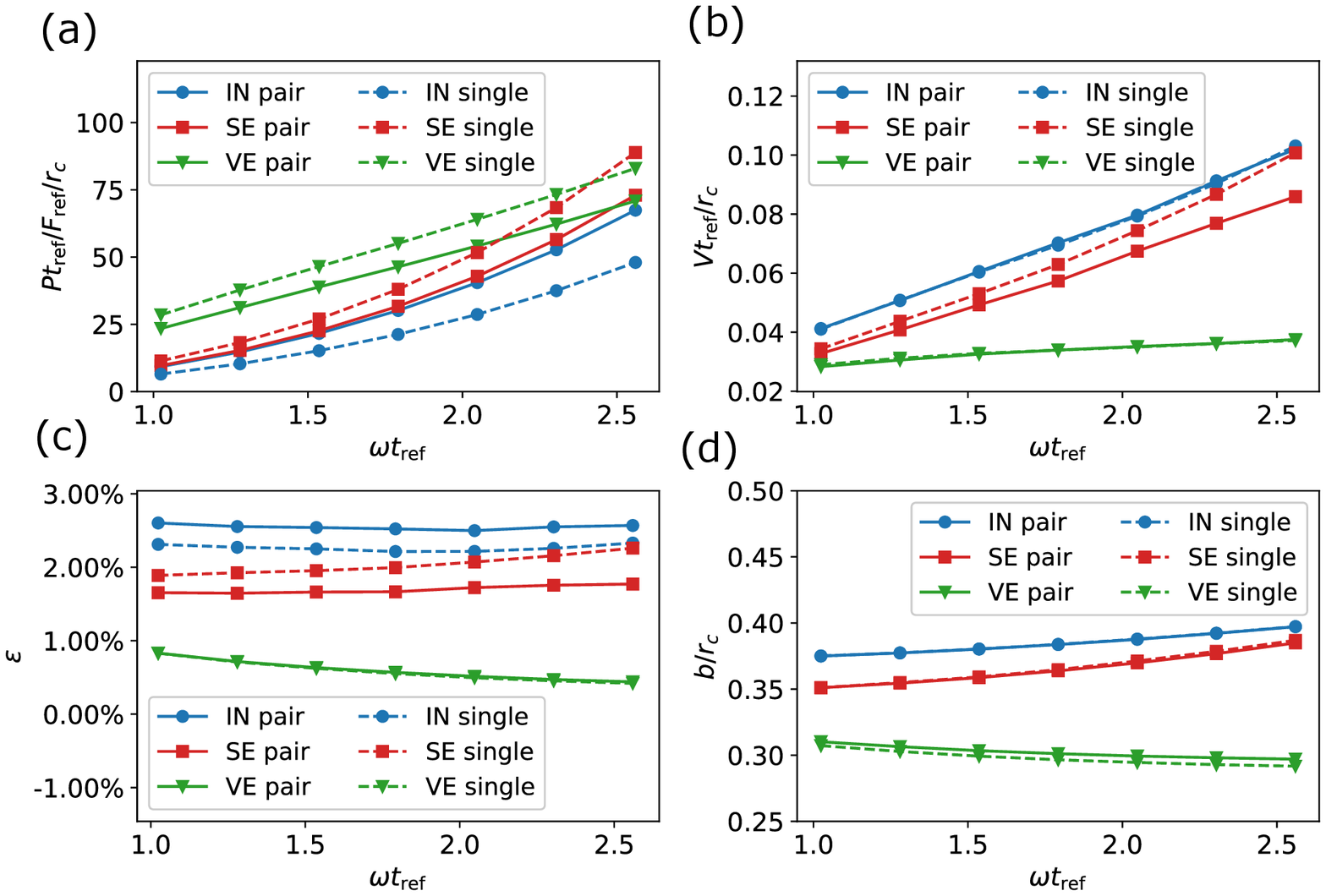}
	\caption{Swimming properties of two synchronized sheets in comparison to a single sheet. (a) Sheet output power $P$, (b) swimming velocity $V$,
		(c) swimming efficiency $\mathcal{E}$, and (d) beating amplitude $b$ for various conditions. The abbreviation ``IN'' denotes Newtonian-fluid simulations, 
		in which inertial effects dominate, with the following parameters $\eta / \eta_\mathrm{ref} = 1.6$, $\rho/\rho_\mathrm{ref}=1$, $\theta_b / \pi=0.044$, 
		$\zeta_\mathrm{s} / \zeta_\mathrm{ref} = 1638.4$ ($\kappa/\kappa_\mathrm{ref}=211.2$), $p_0 / p_\mathrm{ref}=2621.4$, and $\nu=7$. The abbreviation ``SE'' denotes Newtonian-fluid simulations, 
		in which the effect of sheet flexibility dominates, with simulations parameters $\eta / \eta_\mathrm{ref}=3.2$, $\rho/\rho_\mathrm{ref}=1$, $\theta_b / \pi=0.028$, 
		$\zeta_\mathrm{s} / \zeta_\mathrm{ref} = 81.92$ ($\kappa/\kappa_\mathrm{ref}=162.6$), $p_0 / p_\mathrm{ref} =819.2$, and $\nu=7$. ``VE'' corresponds to non-Newtonian-fluid simulations with 
		dominating viscoelastic effects for simulation parameters $\eta_\mathrm{s} / \eta_\mathrm{ref}=3.0$, $\eta_\mathrm{p} / \eta_\mathrm{ref}=13.0$, 
		$\tau / t_\mathrm{ref} = 2.0$, $\rho/\rho_\mathrm{ref}=0.125$, $\theta_b / \pi=0.056$, $\zeta_\mathrm{s} / \zeta_\mathrm{ref} = 4096$ ($\kappa/\kappa_\mathrm{ref}=288$), 
		$p_0 / p_\mathrm{ref} =6553.6$, and $\nu=7$.} 
	\label{fig:13}
\end{figure}

\subsection{Swimming efficiency of two synchronized sheets}
\label{sec:eff}
       
Different mechanisms, such as inertia, sheet elasticity, and fluid compressibility and viscoelasticity, can contribute to the synchronization of 
two sheets. An interesting question is whether the behavior of two synchronized sheets is different from that of a single sheet. 
In the early work of Taylor \cite{Taylor_ASO_1951}, it has been shown that energy dissipation of two sheets at $\mathrm{Re}=0$ 
is minimized (maximized) when they attain the in-phase (opposite-phase) configuration. More recent theoretical study \cite{Elfring_HPL_2009} 
reports that the stable synchronized phase is not necessarily the phase with a minimum energy dissipation. There exist numerous examples 
of biological micro-organisms swimming in clusters, suggesting that collective swimming may have some advantages. A numerical study 
about sperm swimming \cite{Yang_CS2D_2008} reports that clustered sperms swim slower, and consume a lower amount of energy per 
sperm than a single one alone. However, it is not clear whether different factors (e.g., inertia, sheet elasticity, and fluid
viscoelasticity) affect the properties of synchronized sheets in a qualitatively similar way. 
 
Figure \ref{fig:13} compares output power, swimming velocity and efficiency of a pair of synchronized sheets with those of a single 
sheet. In these simulations, two sheets are let to fully synchronize, and after that the aforementioned measurements are performed. 
Three different synchronization factors are considered, including sheet synchronization dominated by inertial effects (denoted as ``IN''), 
sheet elasticity (abbreviated as ``SE''), and fluid viscoelasticity (denoted by ``VE''). Figure \ref{fig:13}(a) presents total output 
power for the three cases, which is computed as $P = -\sum_{i\le N} \sum_{j \le M} \bm{F}_{ij} \cdot \mathbf{v}_i$, where $N$ is the total 
number of sheet particles, $M$ is the total number of fluid particles, and $\bm{F}_{ij}$ are inter-particle forces. Note that only for 
the inertia-dominated case, the output power of a synchronized pair is larger than that of a single sheet. This is due to the fact that
the sheets synchronize to the opposite-phase conformation, for which the dissipation energy is largest in Stokes flow regime 
($\mathrm{Re}$ here is smaller than $0.04$). Interestingly, swimming velocities of a synchronized pair and a single sheet do not differ 
much, see Fig.~\ref{fig:13}(b). Only in the "SE" case, the swimming velocity of a single sheet is slightly larger than that of the pair. 
These results indicate that only in the "VE" case, the synchronized pair of sheets swims not slower than the corresponding single sheet,
and has a lower output power. 

Figure \ref{fig:13}(c) presents swimming efficiency $\mathcal{E} = (P-P_\mathrm{eff})/P$ for different cases, where $P_\mathrm{eff} = 
F_x^\mathrm{visc} V_x$ is the effective power with $F_x^\mathrm{visc}$ being the $x$ component of viscous forces exerted by the 
fluid on the sheets and $V_x$ is the swimming speed. In the case of dominating inertial effects, the swimming efficiency (about $2-3\%$) is
largest, and the synchronized pair is slightly more efficient than a single sheet. In the case when sheet-elasticity effects dominate, 
the pair has a lower efficiency than the single sheet. Finally, in case of dominating viscoelastic contributions, the swimming efficiency 
is smallest and there is no difference in $\mathcal{E}$ for the synchronized pair and single sheet. Nevertheless, swimming efficiency 
may not be an appropriate measure to clearly determine possible advantages/disadvantages of synchronized swimming. For example, for the 
``IN'' case in Fig.~\ref{fig:13}, the efficiency and total output power are larger for the synchronized pair than for the single sheet, 
but the swimming speed is nearly the same. In this case, the sheets synchronize toward the opposite-phase configuration, which 
results in a relatively strong backward (peristaltic-like) flow between them, thus increasing the resistance for swimming. Finally, 
Fig.~\ref{fig:13}(d) shows that wave amplitudes are nearly the same for both the synchronized pair and single sheet in all cases.

\begin{table}
	\centering
	\caption{Different factors which lead to the synchronization of two sheets toward the in-phase or opposite-phase configurations. When several factors are 
	present, the final configuration is determined by their competition.}
	\label{tab:sum}
	\begin{tabular}{p{3.2cm} p{6.8cm} p{6.8cm}}
		\hline	
		\textbf{Factors} & \textbf{Theory} &  \textbf{Simulation}\\
		\hline
		No inertia ($\mathrm{Re}=0$) & No synchronization (incompressible fluid, inextensible sheets) \cite{Elfring_HPL_2009} &  \\
		\hline
		Inertia ($\mathrm{Re} > 0$)  &   & Opposite-phase configuration (this study and Ref.~\cite{Fauci_IOF_1990}) \\
		\hline
		Asymmetric wave & In-phase or opposite-phase configuration, depending on the asymmetry ($\mathrm{Re}=0$) \cite{Elfring_HPL_2009,Elfring_PHS_2011} &   \\ 
		\hline
		Sheet flexibility & In-phase configuration ($\mathrm{Re}=0$) \cite{Elfring_SFS_2011} &  In-phase configuration ($\mathrm{Re} > 0$, this study) \\
		\hline
		Fluid compressibility &   & In-phase configuration ($\mathrm{Re} > 0$, this study)  \\
		\hline
		Fluid viscoelasticity & In-phase configuration \cite{Elfring_2DS_2010} & In-phase configuration (this study and Ref.~\cite{Chrispell_AES_2013}) \\
		\hline
	\end{tabular}
\end{table}

\section{Conclusions}
\label{sec:conc}

We have employed numerical simulations to study the effect of inertia, sheet flexibility, and fluid compressibility and viscoelasticity on the synchronization 
of two inextensible or flexible sheets. Table \ref{tab:sum} shows the summary of all results. Inertial effects always lead to sheet synchronization toward the opposite-phase configuration. 
When inertial effects dominate, the synchronization time $\tau^\mathrm{s}$ of two sheets is inversely proportional to the Reynolds number, 
such that $\tau^\mathrm{s} \omega \propto \mathrm{Re}^{-1}$. Both fluid compressibility and sheet flexibility drive synchronization toward 
the in-phase configuration, and compete with inertial effects for $\mathrm{Re} > 0$. Furthermore, we have systematically tested the theoretical 
prediction \cite{Elfring_2DS_2010} of the synchronization forces between two sheets in viscoelastic fluids, favoring the in-phase configuration.
Our simulation results are in excellent agreement with the theoretical prediction for $bk < 0.4$; however, for large $bk$, synchronization forces strongly
depart from the theory for $\mathrm{De} > 1$, indicating a rapid synchronization. Thus, for $\mathrm{De} > 1$ and large 
enough $bk$, fluid viscoelasticity has a dramatic effect on the synchronization of two sheets, which clearly dominates over other factors,
such as inertia and sheet elasticity. This result is consistent with the observations of significant enhancement of sperm clustering in
viscoelastic fluids \cite{Tung_FVE_2017}. Finally, sheet synchronization dominated by fluid viscoelasticity does not impede swimming
velocity of the synchronized pair, but has a lower output power in comparison to a single sheet.  

Simulation results presented here constitute a systematic study of competing effects for the synchronization of two sheets. They can be 
used to qualitatively assess the importance of possible factors for experimentally-observed interactions between biological microswimmers 
or artificial microrobots. This knowledge is useful for a better understanding of collective behavior of microswimmers and for tuning 
of synchronization interactions between artificial microrobots.

\section*{Acknowledgments}
C.M. acknowledges funding by the China Scholarship Council (CSC) and German Academic Exchange Service (DAAD) through the Sino-German 
(CSC-DAAD) Postdoc Scholarship Program. We gratefully acknowledge the computing time granted through JARA-HPC on the supercomputer 
JURECA \cite{jureca} at Forschungszentrum J\"ulich.

\appendix
\section{Calculation of flexural rigidity $\kappa$}

\begin{figure}[!htb]
     \centering
     \includegraphics[width=.4\linewidth]{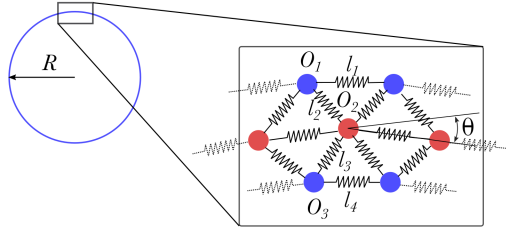}
     \caption{Schematic of a ring used for the calculation of sheet flexural rigidity $\kappa$. $R$ is the ring radius, $l_1, \dots, l_4$ are lengths
     	of the corresponding springs, and $\theta$ is the angle between two adjacent springs in the middle layer. Equilibrium lengths are $l_1^0 = l_4^0 = l_0$
     	and $l_2^0 = l_3^0 = \sqrt{5}l_0/2$.}
     \label{fig:a1}
\end{figure}

We consider a ring (see Fig.~\ref{fig:a1}) whose structure is similar to the sheet in Fig.~\ref{fig:01}(b). In continuum, elastic energy of the ring is
given by
\begin{equation}
	E = \frac{\kappa}{2}\int_{2\pi R} \frac{\mathrm{d}l}{R^2} = \frac{\pi \kappa}{R},
	\label{eq:Ec}
\end{equation}
which can be compared with the energy of a discrete structure. The force balance at $O_1$ yields
\begin{equation}
	2\zeta_\mathrm{s}(l_1-l_0)\sin\frac{\theta}{2} = 2\zeta_\mathrm{s}(\frac{\sqrt{5}}{2}l_0-l_2)\cdot \frac{2}{\sqrt{5}},
\end{equation}
resulting in
\begin{equation}
	l_2 \approx \frac{\sqrt{5}}{2}(l_0-\Delta l \frac{\theta}{2}),
\end{equation}
with $\Delta l=l_1-l_0$. Under the assumption that the middle layer does not deform, the force balance at $O_2$ leads to $l_2=l_3$.
From the discrete geometry in Fig.~\ref{fig:a1}, we obtain
\begin{equation}
	\frac{R}{l_0} = \frac{R+\frac{2}{\sqrt{5}}l_2}{l_0+\Delta l} \quad \Rightarrow \quad \Delta l = \frac{2l_0^2R}{2R^2+l_0^2}.
	\label{eq:deltal}
\end{equation}
Then, the elastic energy of the discrete structure is given by
\begin{equation}
	E_d = N\zeta_\mathrm{s}(\Delta l^2 + \frac{5}{8}\Delta l^2\theta^2) + N\frac{\zeta_\theta}{2}\theta^2.
\end{equation}
By substituting $\Delta l$ and $\theta=l_0/R$ into the equation above, we obtain
\begin{equation}
	E_d\approx \frac{2\pi\zeta_\mathrm{s}l_0^3+\pi\zeta_\theta l_0}{R},
	\label{eq:Ed}
\end{equation}
which results, when compared with Eq.~(\ref{eq:Ec}), in the expression for flexural rigidity $\kappa$ as
\begin{equation}
	\kappa = 2\zeta_\mathrm{s}l_0^3+\zeta_\theta l_0.
\end{equation}


\begin{thebibliography}{50}%
\makeatletter
\providecommand \@ifxundefined [1]{%
 \@ifx{#1\undefined}
}%
\providecommand \@ifnum [1]{%
 \ifnum #1\expandafter \@firstoftwo
 \else \expandafter \@secondoftwo
 \fi
}%
\providecommand \@ifx [1]{%
 \ifx #1\expandafter \@firstoftwo
 \else \expandafter \@secondoftwo
 \fi
}%
\providecommand \natexlab [1]{#1}%
\providecommand \enquote  [1]{``#1''}%
\providecommand \bibnamefont  [1]{#1}%
\providecommand \bibfnamefont [1]{#1}%
\providecommand \citenamefont [1]{#1}%
\providecommand \href@noop [0]{\@secondoftwo}%
\providecommand \href [0]{\begingroup \@sanitize@url \@href}%
\providecommand \@href[1]{\@@startlink{#1}\@@href}%
\providecommand \@@href[1]{\endgroup#1\@@endlink}%
\providecommand \@sanitize@url [0]{\catcode `\\12\catcode `\$12\catcode
  `\&12\catcode `\#12\catcode `\^12\catcode `\_12\catcode `\%12\relax}%
\providecommand \@@startlink[1]{}%
\providecommand \@@endlink[0]{}%
\providecommand \url  [0]{\begingroup\@sanitize@url \@url }%
\providecommand \@url [1]{\endgroup\@href {#1}{\urlprefix }}%
\providecommand \urlprefix  [0]{URL }%
\providecommand \Eprint [0]{\href }%
\providecommand \doibase [0]{http://dx.doi.org/}%
\providecommand \selectlanguage [0]{\@gobble}%
\providecommand \bibinfo  [0]{\@secondoftwo}%
\providecommand \bibfield  [0]{\@secondoftwo}%
\providecommand \translation [1]{[#1]}%
\providecommand \BibitemOpen [0]{}%
\providecommand \bibitemStop [0]{}%
\providecommand \bibitemNoStop [0]{.\EOS\space}%
\providecommand \EOS [0]{\spacefactor3000\relax}%
\providecommand \BibitemShut  [1]{\csname bibitem#1\endcsname}%
\let\auto@bib@innerbib\@empty
\bibitem [{\citenamefont {Lauga}\ and\ \citenamefont
  {Powers}(2009)}]{Lauga_HSO_2009}%
  \BibitemOpen
  \bibfield  {author} {\bibinfo {author} {\bibfnamefont {E.}~\bibnamefont
  {Lauga}}\ and\ \bibinfo {author} {\bibfnamefont {T.~R.}\ \bibnamefont
  {Powers}},\ }\bibfield  {title} {\enquote {\bibinfo {title} {The
  hydrodynamics of swimming microorganisms},}\ }\href@noop {} {\bibfield
  {journal} {\bibinfo  {journal} {Rep. Prog. Phys.}\ }\textbf {\bibinfo
  {volume} {72}},\ \bibinfo {pages} {096601} (\bibinfo {year}
  {2009})}\BibitemShut {NoStop}%
\bibitem [{\citenamefont {Elgeti}\ \emph {et~al.}(2015)\citenamefont {Elgeti},
  \citenamefont {Winkler},\ and\ \citenamefont {Gompper}}]{Elgeti_PMS_2015}%
  \BibitemOpen
  \bibfield  {author} {\bibinfo {author} {\bibfnamefont {J.}~\bibnamefont
  {Elgeti}}, \bibinfo {author} {\bibfnamefont {R.~G.}\ \bibnamefont {Winkler}},
  \ and\ \bibinfo {author} {\bibfnamefont {G.}~\bibnamefont {Gompper}},\
  }\bibfield  {title} {\enquote {\bibinfo {title} {Physics of microswimmers -
  single particle motion and collective behavior: a review},}\ }\href@noop {}
  {\bibfield  {journal} {\bibinfo  {journal} {Rep. Prog. Phys.}\ }\textbf
  {\bibinfo {volume} {78}},\ \bibinfo {pages} {056601} (\bibinfo {year}
  {2015})}\BibitemShut {NoStop}%
\bibitem [{\citenamefont {Bechinger}\ \emph {et~al.}(2016)\citenamefont
  {Bechinger}, \citenamefont {Di~Leonardo}, \citenamefont {L{\"o}wen},
  \citenamefont {Reichhardt}, \citenamefont {Volpe},\ and\ \citenamefont
  {Volpe}}]{Bechinger_APC_2016}%
  \BibitemOpen
  \bibfield  {author} {\bibinfo {author} {\bibfnamefont {C.}~\bibnamefont
  {Bechinger}}, \bibinfo {author} {\bibfnamefont {R.}~\bibnamefont
  {Di~Leonardo}}, \bibinfo {author} {\bibfnamefont {H.}~\bibnamefont
  {L{\"o}wen}}, \bibinfo {author} {\bibfnamefont {C.}~\bibnamefont
  {Reichhardt}}, \bibinfo {author} {\bibfnamefont {G.}~\bibnamefont {Volpe}}, \
  and\ \bibinfo {author} {\bibfnamefont {G.}~\bibnamefont {Volpe}},\ }\bibfield
   {title} {\enquote {\bibinfo {title} {Active particles in complex and crowded
  environments},}\ }\href@noop {} {\bibfield  {journal} {\bibinfo  {journal}
  {Rev. Mod. Phys.}\ }\textbf {\bibinfo {volume} {88}},\ \bibinfo {pages}
  {045006} (\bibinfo {year} {2016})}\BibitemShut {NoStop}%
\bibitem [{\citenamefont {Palagi}\ and\ \citenamefont
  {Fischer}(2018)}]{Palagi_BIR_2018}%
  \BibitemOpen
  \bibfield  {author} {\bibinfo {author} {\bibfnamefont {S.}~\bibnamefont
  {Palagi}}\ and\ \bibinfo {author} {\bibfnamefont {P.}~\bibnamefont
  {Fischer}},\ }\bibfield  {title} {\enquote {\bibinfo {title} {Bioinspired
  microrobots},}\ }\href@noop {} {\bibfield  {journal} {\bibinfo  {journal}
  {Nat. Rev. Mater.}\ }\textbf {\bibinfo {volume} {3}},\ \bibinfo {pages}
  {113--124} (\bibinfo {year} {2018})}\BibitemShut {NoStop}%
\bibitem [{\citenamefont {Woolley}\ \emph {et~al.}(2009)\citenamefont
  {Woolley}, \citenamefont {Crockett}, \citenamefont {Groom},\ and\
  \citenamefont {Revell}}]{Woolley_SSF_2009}%
  \BibitemOpen
  \bibfield  {author} {\bibinfo {author} {\bibfnamefont {D.~M.}\ \bibnamefont
  {Woolley}}, \bibinfo {author} {\bibfnamefont {R.~F.}\ \bibnamefont
  {Crockett}}, \bibinfo {author} {\bibfnamefont {W.~D.~I.}\ \bibnamefont
  {Groom}}, \ and\ \bibinfo {author} {\bibfnamefont {S.~G.}\ \bibnamefont
  {Revell}},\ }\bibfield  {title} {\enquote {\bibinfo {title} {A study of
  synchronisation between the flagella of bull spermatozoa, with related
  observations},}\ }\href@noop {} {\bibfield  {journal} {\bibinfo  {journal}
  {J. Exp. Biol.}\ }\textbf {\bibinfo {volume} {212}},\ \bibinfo {pages}
  {2215--2223} (\bibinfo {year} {2009})}\BibitemShut {NoStop}%
\bibitem [{\citenamefont {Yang}\ \emph {et~al.}(2008)\citenamefont {Yang},
  \citenamefont {Elgeti},\ and\ \citenamefont {Gompper}}]{Yang_CS2D_2008}%
  \BibitemOpen
  \bibfield  {author} {\bibinfo {author} {\bibfnamefont {Y.}~\bibnamefont
  {Yang}}, \bibinfo {author} {\bibfnamefont {J.}~\bibnamefont {Elgeti}}, \ and\
  \bibinfo {author} {\bibfnamefont {G.}~\bibnamefont {Gompper}},\ }\bibfield
  {title} {\enquote {\bibinfo {title} {Cooperation of sperm in two dimensions:
  synchronization, attraction, and aggregation through hydrodynamic
  interactions},}\ }\href@noop {} {\bibfield  {journal} {\bibinfo  {journal}
  {Phys. Rev. E}\ }\textbf {\bibinfo {volume} {78}},\ \bibinfo {pages} {061903}
  (\bibinfo {year} {2008})}\BibitemShut {NoStop}%
\bibitem [{\citenamefont {Nosrati}\ \emph {et~al.}(2015)\citenamefont
  {Nosrati}, \citenamefont {Driouchi}, \citenamefont {Yip},\ and\ \citenamefont
  {Sinton}}]{Nosrati_2DS_2015}%
  \BibitemOpen
  \bibfield  {author} {\bibinfo {author} {\bibfnamefont {R.}~\bibnamefont
  {Nosrati}}, \bibinfo {author} {\bibfnamefont {A.}~\bibnamefont {Driouchi}},
  \bibinfo {author} {\bibfnamefont {C.~M.}\ \bibnamefont {Yip}}, \ and\
  \bibinfo {author} {\bibfnamefont {D.}~\bibnamefont {Sinton}},\ }\bibfield
  {title} {\enquote {\bibinfo {title} {Two-dimensional slither swimming of
  sperm within a micrometre of a surface},}\ }\href@noop {} {\bibfield
  {journal} {\bibinfo  {journal} {Nat. Comm.}\ }\textbf {\bibinfo {volume}
  {6}},\ \bibinfo {pages} {8703} (\bibinfo {year} {2015})}\BibitemShut
  {NoStop}%
\bibitem [{\citenamefont {Kim}\ \emph {et~al.}(2003)\citenamefont {Kim},
  \citenamefont {Bird}, \citenamefont {Van~Parys}, \citenamefont {Breuer},\
  and\ \citenamefont {Powers}}]{Kim_MSM_2003}%
  \BibitemOpen
  \bibfield  {author} {\bibinfo {author} {\bibfnamefont {M.-J.}\ \bibnamefont
  {Kim}}, \bibinfo {author} {\bibfnamefont {J.~C.}\ \bibnamefont {Bird}},
  \bibinfo {author} {\bibfnamefont {A.~J.}\ \bibnamefont {Van~Parys}}, \bibinfo
  {author} {\bibfnamefont {K.~S.}\ \bibnamefont {Breuer}}, \ and\ \bibinfo
  {author} {\bibfnamefont {T.~R.}\ \bibnamefont {Powers}},\ }\bibfield  {title}
  {\enquote {\bibinfo {title} {A macroscopic scale model of bacterial flagellar
  bundling},}\ }\href@noop {} {\bibfield  {journal} {\bibinfo  {journal} {Proc.
  Natl. Acad. Sci. USA}\ }\textbf {\bibinfo {volume} {100}},\ \bibinfo {pages}
  {15481--15485} (\bibinfo {year} {2003})}\BibitemShut {NoStop}%
\bibitem [{\citenamefont {Reigh}\ \emph {et~al.}(2012)\citenamefont {Reigh},
  \citenamefont {Winkler},\ and\ \citenamefont {Gompper}}]{Reigh_SBF_2012}%
  \BibitemOpen
  \bibfield  {author} {\bibinfo {author} {\bibfnamefont {S.~Y.}\ \bibnamefont
  {Reigh}}, \bibinfo {author} {\bibfnamefont {R.~G.}\ \bibnamefont {Winkler}},
  \ and\ \bibinfo {author} {\bibfnamefont {G.}~\bibnamefont {Gompper}},\
  }\bibfield  {title} {\enquote {\bibinfo {title} {Synchronization and bundling
  of anchored bacterial flagella},}\ }\href@noop {} {\bibfield  {journal}
  {\bibinfo  {journal} {Soft Matter}\ }\textbf {\bibinfo {volume} {8}},\
  \bibinfo {pages} {4363--4372} (\bibinfo {year} {2012})}\BibitemShut {NoStop}%
\bibitem [{\citenamefont {Brumley}\ \emph {et~al.}(2012)\citenamefont
  {Brumley}, \citenamefont {Polin}, \citenamefont {Pedley},\ and\ \citenamefont
  {Goldstein}}]{Brumley_HSW_2012}%
  \BibitemOpen
  \bibfield  {author} {\bibinfo {author} {\bibfnamefont {D.~R.}\ \bibnamefont
  {Brumley}}, \bibinfo {author} {\bibfnamefont {M.}~\bibnamefont {Polin}},
  \bibinfo {author} {\bibfnamefont {T.~J.}\ \bibnamefont {Pedley}}, \ and\
  \bibinfo {author} {\bibfnamefont {R.~E.}\ \bibnamefont {Goldstein}},\
  }\bibfield  {title} {\enquote {\bibinfo {title} {Hydrodynamic synchronization
  and metachronal waves on the surface of the colonial alga {V}olvox
  carteri},}\ }\href@noop {} {\bibfield  {journal} {\bibinfo  {journal} {Phys.
  Rev. Lett.}\ }\textbf {\bibinfo {volume} {109}},\ \bibinfo {pages} {268102}
  (\bibinfo {year} {2012})}\BibitemShut {NoStop}%
\bibitem [{\citenamefont {Elgeti}\ and\ \citenamefont
  {Gompper}(2013)}]{Elgeti_EMW_2013}%
  \BibitemOpen
  \bibfield  {author} {\bibinfo {author} {\bibfnamefont {J.}~\bibnamefont
  {Elgeti}}\ and\ \bibinfo {author} {\bibfnamefont {G.}~\bibnamefont
  {Gompper}},\ }\bibfield  {title} {\enquote {\bibinfo {title} {Emergence of
  metachronal waves in cilia arrays},}\ }\href@noop {} {\bibfield  {journal}
  {\bibinfo  {journal} {Proc. Natl. Acad. Sci. USA}\ }\textbf {\bibinfo
  {volume} {110}},\ \bibinfo {pages} {4470--4475} (\bibinfo {year}
  {2013})}\BibitemShut {NoStop}%
\bibitem [{\citenamefont {Taylor}(1951)}]{Taylor_ASO_1951}%
  \BibitemOpen
  \bibfield  {author} {\bibinfo {author} {\bibfnamefont {G.~I.}\ \bibnamefont
  {Taylor}},\ }\bibfield  {title} {\enquote {\bibinfo {title} {Analysis of the
  swimming of microscopic organisms},}\ }\href@noop {} {\bibfield  {journal}
  {\bibinfo  {journal} {Proc. R. Soc. Lond. A}\ }\textbf {\bibinfo {volume}
  {209}},\ \bibinfo {pages} {447--461} (\bibinfo {year} {1951})}\BibitemShut
  {NoStop}%
\bibitem [{\citenamefont {Brumley}\ \emph {et~al.}(2014)\citenamefont
  {Brumley}, \citenamefont {Wan}, \citenamefont {Polin},\ and\ \citenamefont
  {Goldstein}}]{Brumley_FSH_2014}%
  \BibitemOpen
  \bibfield  {author} {\bibinfo {author} {\bibfnamefont {D.~R.}\ \bibnamefont
  {Brumley}}, \bibinfo {author} {\bibfnamefont {K.~Y.}\ \bibnamefont {Wan}},
  \bibinfo {author} {\bibfnamefont {M.}~\bibnamefont {Polin}}, \ and\ \bibinfo
  {author} {\bibfnamefont {R.~E.}\ \bibnamefont {Goldstein}},\ }\bibfield
  {title} {\enquote {\bibinfo {title} {Flagellar synchronization through direct
  hydrodynamic interactions},}\ }\href@noop {} {\bibfield  {journal} {\bibinfo
  {journal} {eLife}\ }\textbf {\bibinfo {volume} {3}},\ \bibinfo {pages}
  {e02750} (\bibinfo {year} {2014})}\BibitemShut {NoStop}%
\bibitem [{\citenamefont {Pedley}\ and\ \citenamefont
  {Kessler}(1992)}]{Pedley_HPS_1992}%
  \BibitemOpen
  \bibfield  {author} {\bibinfo {author} {\bibfnamefont {T.~J.}\ \bibnamefont
  {Pedley}}\ and\ \bibinfo {author} {\bibfnamefont {J.~O.}\ \bibnamefont
  {Kessler}},\ }\bibfield  {title} {\enquote {\bibinfo {title} {Hydrodynamic
  phenomena in suspensions of swimming microorganisms},}\ }\href@noop {}
  {\bibfield  {journal} {\bibinfo  {journal} {Annu. Rev. Fluid Mech.}\ }\textbf
  {\bibinfo {volume} {24}},\ \bibinfo {pages} {313--358} (\bibinfo {year}
  {1992})}\BibitemShut {NoStop}%
\bibitem [{\citenamefont {Golestanian}\ \emph {et~al.}(2011)\citenamefont
  {Golestanian}, \citenamefont {Yeomans},\ and\ \citenamefont
  {Uchida}}]{Golestanian_HSR_2011}%
  \BibitemOpen
  \bibfield  {author} {\bibinfo {author} {\bibfnamefont {R.}~\bibnamefont
  {Golestanian}}, \bibinfo {author} {\bibfnamefont {J.~M.}\ \bibnamefont
  {Yeomans}}, \ and\ \bibinfo {author} {\bibfnamefont {N.}~\bibnamefont
  {Uchida}},\ }\bibfield  {title} {\enquote {\bibinfo {title} {Hydrodynamic
  synchronization at low {R}eynolds number},}\ }\href@noop {} {\bibfield
  {journal} {\bibinfo  {journal} {Soft Matter}\ }\textbf {\bibinfo {volume}
  {7}},\ \bibinfo {pages} {3074--3082} (\bibinfo {year} {2011})}\BibitemShut
  {NoStop}%
\bibitem [{\citenamefont {Reynolds}(1965)}]{Reynolds_SMO_1965}%
  \BibitemOpen
  \bibfield  {author} {\bibinfo {author} {\bibfnamefont {A.~J.}\ \bibnamefont
  {Reynolds}},\ }\bibfield  {title} {\enquote {\bibinfo {title} {The swimming
  of minute organisms},}\ }\href@noop {} {\bibfield  {journal} {\bibinfo
  {journal} {J. Fluid Mech.}\ }\textbf {\bibinfo {volume} {23}},\ \bibinfo
  {pages} {241--260} (\bibinfo {year} {1965})}\BibitemShut {NoStop}%
\bibitem [{\citenamefont {Tuck}(1968)}]{Tuck_NSP_1968}%
  \BibitemOpen
  \bibfield  {author} {\bibinfo {author} {\bibfnamefont {E.~O.}\ \bibnamefont
  {Tuck}},\ }\bibfield  {title} {\enquote {\bibinfo {title} {A note on a
  swimming problem},}\ }\href@noop {} {\bibfield  {journal} {\bibinfo
  {journal} {J. Fluid Mech.}\ }\textbf {\bibinfo {volume} {31}},\ \bibinfo
  {pages} {305--308} (\bibinfo {year} {1968})}\BibitemShut {NoStop}%
\bibitem [{\citenamefont {Diller}\ \emph {et~al.}(2014)\citenamefont {Diller},
  \citenamefont {Zhuang}, \citenamefont {Lum}, \citenamefont {Edwards},\ and\
  \citenamefont {Sitti}}]{Diller_CDM_2014}%
  \BibitemOpen
  \bibfield  {author} {\bibinfo {author} {\bibfnamefont {E.}~\bibnamefont
  {Diller}}, \bibinfo {author} {\bibfnamefont {J.}~\bibnamefont {Zhuang}},
  \bibinfo {author} {\bibfnamefont {G.~Z.}\ \bibnamefont {Lum}}, \bibinfo
  {author} {\bibfnamefont {M.~R.}\ \bibnamefont {Edwards}}, \ and\ \bibinfo
  {author} {\bibfnamefont {M.}~\bibnamefont {Sitti}},\ }\bibfield  {title}
  {\enquote {\bibinfo {title} {Continuously distributed magnetization profile
  for millimeter-scale elastomeric undulatory swimming},}\ }\href@noop {}
  {\bibfield  {journal} {\bibinfo  {journal} {Appl. Phys. Lett.}\ }\textbf
  {\bibinfo {volume} {104}},\ \bibinfo {pages} {174101} (\bibinfo {year}
  {2014})}\BibitemShut {NoStop}%
\bibitem [{\citenamefont {Huang}\ \emph {et~al.}(2016)\citenamefont {Huang},
  \citenamefont {Sakar}, \citenamefont {Petruska}, \citenamefont {Pan{\'e}},\
  and\ \citenamefont {Nelson}}]{Huang_SMM_2016}%
  \BibitemOpen
  \bibfield  {author} {\bibinfo {author} {\bibfnamefont {H.-W.}\ \bibnamefont
  {Huang}}, \bibinfo {author} {\bibfnamefont {M.~S.}\ \bibnamefont {Sakar}},
  \bibinfo {author} {\bibfnamefont {A.~J.}\ \bibnamefont {Petruska}}, \bibinfo
  {author} {\bibfnamefont {S.}~\bibnamefont {Pan{\'e}}}, \ and\ \bibinfo
  {author} {\bibfnamefont {B.~J.}\ \bibnamefont {Nelson}},\ }\bibfield  {title}
  {\enquote {\bibinfo {title} {Soft micromachines with programmable motility
  and morphology},}\ }\href@noop {} {\bibfield  {journal} {\bibinfo  {journal}
  {Nat. Comm.}\ }\textbf {\bibinfo {volume} {7}},\ \bibinfo {pages} {12263}
  (\bibinfo {year} {2016})}\BibitemShut {NoStop}%
\bibitem [{\citenamefont {Elfring}\ and\ \citenamefont
  {Lauga}(2009)}]{Elfring_HPL_2009}%
  \BibitemOpen
  \bibfield  {author} {\bibinfo {author} {\bibfnamefont {G.~J.}\ \bibnamefont
  {Elfring}}\ and\ \bibinfo {author} {\bibfnamefont {E.}~\bibnamefont
  {Lauga}},\ }\bibfield  {title} {\enquote {\bibinfo {title} {Hydrodynamic
  phase locking of swimming microorganisms},}\ }\href@noop {} {\bibfield
  {journal} {\bibinfo  {journal} {Phys. Rev. Lett.}\ }\textbf {\bibinfo
  {volume} {103}},\ \bibinfo {pages} {088101} (\bibinfo {year}
  {2009})}\BibitemShut {NoStop}%
\bibitem [{\citenamefont {Friedrich}(2016)}]{Friedrich_HSO_2016}%
  \BibitemOpen
  \bibfield  {author} {\bibinfo {author} {\bibfnamefont {B.}~\bibnamefont
  {Friedrich}},\ }\bibfield  {title} {\enquote {\bibinfo {title} {Hydrodynamic
  synchronization of flagellar oscillators},}\ }\href@noop {} {\bibfield
  {journal} {\bibinfo  {journal} {Eur. Phys. J. Special Topics}\ }\textbf
  {\bibinfo {volume} {225}},\ \bibinfo {pages} {2353--2368} (\bibinfo {year}
  {2016})}\BibitemShut {NoStop}%
\bibitem [{\citenamefont {Elfring}\ and\ \citenamefont
  {Lauga}(2011{\natexlab{a}})}]{Elfring_PHS_2011}%
  \BibitemOpen
  \bibfield  {author} {\bibinfo {author} {\bibfnamefont {G.~J.}\ \bibnamefont
  {Elfring}}\ and\ \bibinfo {author} {\bibfnamefont {E.}~\bibnamefont
  {Lauga}},\ }\bibfield  {title} {\enquote {\bibinfo {title} {Passive
  hydrodynamic synchronization of two-dimensional swimming cells},}\
  }\href@noop {} {\bibfield  {journal} {\bibinfo  {journal} {Phys. Fluids}\
  }\textbf {\bibinfo {volume} {23}},\ \bibinfo {pages} {011902} (\bibinfo
  {year} {2011}{\natexlab{a}})}\BibitemShut {NoStop}%
\bibitem [{\citenamefont {Fauci}(1990)}]{Fauci_IOF_1990}%
  \BibitemOpen
  \bibfield  {author} {\bibinfo {author} {\bibfnamefont {L.~J.}\ \bibnamefont
  {Fauci}},\ }\bibfield  {title} {\enquote {\bibinfo {title} {Interaction of
  oscillating filaments: a computational study},}\ }\href@noop {} {\bibfield
  {journal} {\bibinfo  {journal} {J. Comp. Phys.}\ }\textbf {\bibinfo {volume}
  {86}},\ \bibinfo {pages} {294--313} (\bibinfo {year} {1990})}\BibitemShut
  {NoStop}%
\bibitem [{\citenamefont {Fauci}\ and\ \citenamefont
  {McDonald}(1995)}]{Fauci_SMB_1995}%
  \BibitemOpen
  \bibfield  {author} {\bibinfo {author} {\bibfnamefont {L.~J.}\ \bibnamefont
  {Fauci}}\ and\ \bibinfo {author} {\bibfnamefont {A.}~\bibnamefont
  {McDonald}},\ }\bibfield  {title} {\enquote {\bibinfo {title} {Sperm motility
  in the presence of boundaries},}\ }\href@noop {} {\bibfield  {journal}
  {\bibinfo  {journal} {Bull. Math. Biol.}\ }\textbf {\bibinfo {volume} {57}},\
  \bibinfo {pages} {679--699} (\bibinfo {year} {1995})}\BibitemShut {NoStop}%
\bibitem [{\citenamefont {Theers}\ and\ \citenamefont
  {Winkler}(2013)}]{Theers_SRM_2013}%
  \BibitemOpen
  \bibfield  {author} {\bibinfo {author} {\bibfnamefont {M.}~\bibnamefont
  {Theers}}\ and\ \bibinfo {author} {\bibfnamefont {R.~G.}\ \bibnamefont
  {Winkler}},\ }\bibfield  {title} {\enquote {\bibinfo {title} {Synchronization
  of rigid microrotors by time-dependent hydrodynamic interactions},}\
  }\href@noop {} {\bibfield  {journal} {\bibinfo  {journal} {Phys. Rev. E}\
  }\textbf {\bibinfo {volume} {88}},\ \bibinfo {pages} {023012} (\bibinfo
  {year} {2013})}\BibitemShut {NoStop}%
\bibitem [{\citenamefont {Elfring}\ and\ \citenamefont
  {Lauga}(2011{\natexlab{b}})}]{Elfring_SFS_2011}%
  \BibitemOpen
  \bibfield  {author} {\bibinfo {author} {\bibfnamefont {G.~J.}\ \bibnamefont
  {Elfring}}\ and\ \bibinfo {author} {\bibfnamefont {E.}~\bibnamefont
  {Lauga}},\ }\bibfield  {title} {\enquote {\bibinfo {title} {Synchronization
  of flexible sheets},}\ }\href@noop {} {\bibfield  {journal} {\bibinfo
  {journal} {J. Fluid Mech.}\ }\textbf {\bibinfo {volume} {674}},\ \bibinfo
  {pages} {163--173} (\bibinfo {year} {2011}{\natexlab{b}})}\BibitemShut
  {NoStop}%
\bibitem [{\citenamefont {Elfring}\ \emph {et~al.}(2010)\citenamefont
  {Elfring}, \citenamefont {Pak},\ and\ \citenamefont
  {Lauga}}]{Elfring_2DS_2010}%
  \BibitemOpen
  \bibfield  {author} {\bibinfo {author} {\bibfnamefont {G.~J.}\ \bibnamefont
  {Elfring}}, \bibinfo {author} {\bibfnamefont {O.~S.}\ \bibnamefont {Pak}}, \
  and\ \bibinfo {author} {\bibfnamefont {E.}~\bibnamefont {Lauga}},\ }\bibfield
   {title} {\enquote {\bibinfo {title} {Two-dimensional flagellar
  synchronization in viscoelastic fluids},}\ }\href@noop {} {\bibfield
  {journal} {\bibinfo  {journal} {J. Fluid Mech.}\ }\textbf {\bibinfo {volume}
  {646}},\ \bibinfo {pages} {505--515} (\bibinfo {year} {2010})}\BibitemShut
  {NoStop}%
\bibitem [{\citenamefont {Chrispell}\ \emph {et~al.}(2013)\citenamefont
  {Chrispell}, \citenamefont {Fauci},\ and\ \citenamefont
  {Shelley}}]{Chrispell_AES_2013}%
  \BibitemOpen
  \bibfield  {author} {\bibinfo {author} {\bibfnamefont {J.~C.}\ \bibnamefont
  {Chrispell}}, \bibinfo {author} {\bibfnamefont {L.~J.}\ \bibnamefont
  {Fauci}}, \ and\ \bibinfo {author} {\bibfnamefont {M.}~\bibnamefont
  {Shelley}},\ }\bibfield  {title} {\enquote {\bibinfo {title} {An actuated
  elastic sheet interacting with passive and active structures in a
  viscoelastic fluid},}\ }\href@noop {} {\bibfield  {journal} {\bibinfo
  {journal} {Phys. Fluids}\ }\textbf {\bibinfo {volume} {25}},\ \bibinfo
  {pages} {013103} (\bibinfo {year} {2013})}\BibitemShut {NoStop}%
\bibitem [{\citenamefont {Tung}\ \emph {et~al.}(2017)\citenamefont {Tung},
  \citenamefont {Lin}, \citenamefont {Harvey}, \citenamefont {Fiore},
  \citenamefont {Ardon}, \citenamefont {Wu},\ and\ \citenamefont
  {Suarez}}]{Tung_FVE_2017}%
  \BibitemOpen
  \bibfield  {author} {\bibinfo {author} {\bibfnamefont {C.-K.}\ \bibnamefont
  {Tung}}, \bibinfo {author} {\bibfnamefont {C.}~\bibnamefont {Lin}}, \bibinfo
  {author} {\bibfnamefont {B.}~\bibnamefont {Harvey}}, \bibinfo {author}
  {\bibfnamefont {A.~G.}\ \bibnamefont {Fiore}}, \bibinfo {author}
  {\bibfnamefont {F.}~\bibnamefont {Ardon}}, \bibinfo {author} {\bibfnamefont
  {M.}~\bibnamefont {Wu}}, \ and\ \bibinfo {author} {\bibfnamefont {S.~S.}\
  \bibnamefont {Suarez}},\ }\bibfield  {title} {\enquote {\bibinfo {title}
  {Fluid viscoelasticity promotes collective swimming of sperm},}\ }\href@noop
  {} {\bibfield  {journal} {\bibinfo  {journal} {Sci. Rep.}\ }\textbf {\bibinfo
  {volume} {7}},\ \bibinfo {pages} {3152} (\bibinfo {year} {2017})}\BibitemShut
  {NoStop}%
\bibitem [{\citenamefont {Espa\~{n}ol}\ and\ \citenamefont
  {Revenga}(2003)}]{Espanol_SDPD_2003}%
  \BibitemOpen
  \bibfield  {author} {\bibinfo {author} {\bibfnamefont {P.}~\bibnamefont
  {Espa\~{n}ol}}\ and\ \bibinfo {author} {\bibfnamefont {M.}~\bibnamefont
  {Revenga}},\ }\bibfield  {title} {\enquote {\bibinfo {title} {Smoothed
  dissipative particle dynamics},}\ }\href@noop {} {\bibfield  {journal}
  {\bibinfo  {journal} {Phys. Rev. E}\ }\textbf {\bibinfo {volume} {67}},\
  \bibinfo {pages} {026705} (\bibinfo {year} {2003})}\BibitemShut {NoStop}%
\bibitem [{\citenamefont {M{\"u}ller}\ \emph {et~al.}(2015)\citenamefont
  {M{\"u}ller}, \citenamefont {Fedosov},\ and\ \citenamefont
  {Gompper}}]{Mueller_SDPD_2015}%
  \BibitemOpen
  \bibfield  {author} {\bibinfo {author} {\bibfnamefont {K.}~\bibnamefont
  {M{\"u}ller}}, \bibinfo {author} {\bibfnamefont {D.~A.}\ \bibnamefont
  {Fedosov}}, \ and\ \bibinfo {author} {\bibfnamefont {G.}~\bibnamefont
  {Gompper}},\ }\bibfield  {title} {\enquote {\bibinfo {title} {Smoothed
  dissipative particle dynamics with angular momentum conservation},}\
  }\href@noop {} {\bibfield  {journal} {\bibinfo  {journal} {J. Comp. Phys.}\
  }\textbf {\bibinfo {volume} {281}},\ \bibinfo {pages} {301--315} (\bibinfo
  {year} {2015})}\BibitemShut {NoStop}%
\bibitem [{\citenamefont {V{\'a}zquez-Quesada}\ \emph
  {et~al.}(2009)\citenamefont {V{\'a}zquez-Quesada}, \citenamefont {Ellero},\
  and\ \citenamefont {Espa{\~n}ol}}]{Quesada_VEF_2009}%
  \BibitemOpen
  \bibfield  {author} {\bibinfo {author} {\bibfnamefont {A.}~\bibnamefont
  {V{\'a}zquez-Quesada}}, \bibinfo {author} {\bibfnamefont {M.}~\bibnamefont
  {Ellero}}, \ and\ \bibinfo {author} {\bibfnamefont {P.}~\bibnamefont
  {Espa{\~n}ol}},\ }\bibfield  {title} {\enquote {\bibinfo {title} {Smoothed
  particle hydrodynamic model for viscoelastic fluids with thermal
  fluctuations},}\ }\href@noop {} {\bibfield  {journal} {\bibinfo  {journal}
  {Phys. Rev. E}\ }\textbf {\bibinfo {volume} {79}},\ \bibinfo {pages} {056707}
  (\bibinfo {year} {2009})}\BibitemShut {NoStop}%
\bibitem [{\citenamefont {Monaghan}(1992)}]{Monaghan_SPH_1992}%
  \BibitemOpen
  \bibfield  {author} {\bibinfo {author} {\bibfnamefont {J.~J.}\ \bibnamefont
  {Monaghan}},\ }\bibfield  {title} {\enquote {\bibinfo {title} {Smoothed
  particle hydrodynamics},}\ }\href@noop {} {\bibfield  {journal} {\bibinfo
  {journal} {Annu. Rev. Astron. Astrophys.}\ }\textbf {\bibinfo {volume}
  {30}},\ \bibinfo {pages} {543--574} (\bibinfo {year} {1992})}\BibitemShut
  {NoStop}%
\bibitem [{\citenamefont {Hoogerbrugge}\ and\ \citenamefont
  {Koelman}(1992)}]{Hoogerbrugge_SMH_1992}%
  \BibitemOpen
  \bibfield  {author} {\bibinfo {author} {\bibfnamefont {P.~J.}\ \bibnamefont
  {Hoogerbrugge}}\ and\ \bibinfo {author} {\bibfnamefont {J.~M. V.~A.}\
  \bibnamefont {Koelman}},\ }\bibfield  {title} {\enquote {\bibinfo {title}
  {Simulating microscopic hydrodynamic phenomena with dissipative particle
  dynamics},}\ }\href@noop {} {\bibfield  {journal} {\bibinfo  {journal}
  {Europhys. Lett.}\ }\textbf {\bibinfo {volume} {19}},\ \bibinfo {pages}
  {155--160} (\bibinfo {year} {1992})}\BibitemShut {NoStop}%
\bibitem [{\citenamefont {Espa\~{n}ol}\ and\ \citenamefont
  {Warren}(1995)}]{Espanol_SMO_1995}%
  \BibitemOpen
  \bibfield  {author} {\bibinfo {author} {\bibfnamefont {P.}~\bibnamefont
  {Espa\~{n}ol}}\ and\ \bibinfo {author} {\bibfnamefont {P.}~\bibnamefont
  {Warren}},\ }\bibfield  {title} {\enquote {\bibinfo {title} {Statistical
  mechanics of dissipative particle dynamics},}\ }\href@noop {} {\bibfield
  {journal} {\bibinfo  {journal} {Europhys. Lett.}\ }\textbf {\bibinfo {volume}
  {30}},\ \bibinfo {pages} {191--196} (\bibinfo {year} {1995})}\BibitemShut
  {NoStop}%
\bibitem [{\citenamefont {Hu}\ and\ \citenamefont {Adams}(2006)}]{Hu_AMC_2006}%
  \BibitemOpen
  \bibfield  {author} {\bibinfo {author} {\bibfnamefont {X.~Y.}\ \bibnamefont
  {Hu}}\ and\ \bibinfo {author} {\bibfnamefont {N.~A.}\ \bibnamefont {Adams}},\
  }\bibfield  {title} {\enquote {\bibinfo {title} {Angular-momentum
  conservative smoothed particle dynamics for incompressible viscous flows},}\
  }\href@noop {} {\bibfield  {journal} {\bibinfo  {journal} {Phys. Fluids}\
  }\textbf {\bibinfo {volume} {18}},\ \bibinfo {pages} {101702} (\bibinfo
  {year} {2006})}\BibitemShut {NoStop}%
\bibitem [{\citenamefont {G{\"o}tze}\ \emph {et~al.}(2007)\citenamefont
  {G{\"o}tze}, \citenamefont {Noguchi},\ and\ \citenamefont
  {Gompper}}]{Gotze_RAM_2007}%
  \BibitemOpen
  \bibfield  {author} {\bibinfo {author} {\bibfnamefont {I.~O.}\ \bibnamefont
  {G{\"o}tze}}, \bibinfo {author} {\bibfnamefont {H.}~\bibnamefont {Noguchi}},
  \ and\ \bibinfo {author} {\bibfnamefont {G.}~\bibnamefont {Gompper}},\
  }\bibfield  {title} {\enquote {\bibinfo {title} {Relevance of angular
  momentum conservation in mesoscale hydrodynamics simulations},}\ }\href@noop
  {} {\bibfield  {journal} {\bibinfo  {journal} {Phys. Rev. E}\ }\textbf
  {\bibinfo {volume} {76}},\ \bibinfo {pages} {046705} (\bibinfo {year}
  {2007})}\BibitemShut {NoStop}%
\bibitem [{\citenamefont {Allen}\ and\ \citenamefont
  {Tildesley}(1991)}]{Allen_CSL_1991}%
  \BibitemOpen
  \bibfield  {author} {\bibinfo {author} {\bibfnamefont {M.~P.}\ \bibnamefont
  {Allen}}\ and\ \bibinfo {author} {\bibfnamefont {D.~J.}\ \bibnamefont
  {Tildesley}},\ }\href@noop {} {\emph {\bibinfo {title} {Computer simulation
  of liquids}}}\ (\bibinfo  {publisher} {Clarendon Press},\ \bibinfo {address}
  {New York},\ \bibinfo {year} {1991})\BibitemShut {NoStop}%
\bibitem [{\citenamefont {Lucy}(1977)}]{Lucy_NAT_1977}%
  \BibitemOpen
  \bibfield  {author} {\bibinfo {author} {\bibfnamefont {L.~B.}\ \bibnamefont
  {Lucy}},\ }\bibfield  {title} {\enquote {\bibinfo {title} {A numerical
  approach to the testing the fission hypothesis},}\ }\href@noop {} {\bibfield
  {journal} {\bibinfo  {journal} {Astronom. J.}\ }\textbf {\bibinfo {volume}
  {82}},\ \bibinfo {pages} {1013--1024} (\bibinfo {year} {1977})}\BibitemShut
  {NoStop}%
\bibitem [{\citenamefont {Morris}\ \emph {et~al.}(1997)\citenamefont {Morris},
  \citenamefont {Fox},\ and\ \citenamefont {Zhu}}]{Morris_MLR_1997}%
  \BibitemOpen
  \bibfield  {author} {\bibinfo {author} {\bibfnamefont {J.~P.}\ \bibnamefont
  {Morris}}, \bibinfo {author} {\bibfnamefont {P.~J.}\ \bibnamefont {Fox}}, \
  and\ \bibinfo {author} {\bibfnamefont {Y.}~\bibnamefont {Zhu}},\ }\bibfield
  {title} {\enquote {\bibinfo {title} {Modeling low {R}eynolds number
  incompressible flows using {SPH}},}\ }\href@noop {} {\bibfield  {journal}
  {\bibinfo  {journal} {J. Comp. Phys.}\ }\textbf {\bibinfo {volume} {136}},\
  \bibinfo {pages} {214--226} (\bibinfo {year} {1997})}\BibitemShut {NoStop}%
\bibitem [{\citenamefont {Sigalotti}\ \emph {et~al.}(2003)\citenamefont
  {Sigalotti}, \citenamefont {Klapp}, \citenamefont {Sira}, \citenamefont
  {Mele{\'a}n},\ and\ \citenamefont {Hasmy}}]{Sigalotti_SPH_2003}%
  \BibitemOpen
  \bibfield  {author} {\bibinfo {author} {\bibfnamefont {L.~D.~G.}\
  \bibnamefont {Sigalotti}}, \bibinfo {author} {\bibfnamefont {J.}~\bibnamefont
  {Klapp}}, \bibinfo {author} {\bibfnamefont {E.}~\bibnamefont {Sira}},
  \bibinfo {author} {\bibfnamefont {Y.}~\bibnamefont {Mele{\'a}n}}, \ and\
  \bibinfo {author} {\bibfnamefont {A.}~\bibnamefont {Hasmy}},\ }\bibfield
  {title} {\enquote {\bibinfo {title} {{SPH} simulations of time-dependent
  {P}oiseuille flow at low {R}eynolds numbers},}\ }\href@noop {} {\bibfield
  {journal} {\bibinfo  {journal} {J. Comp. Phys.}\ }\textbf {\bibinfo {volume}
  {191}},\ \bibinfo {pages} {622--638} (\bibinfo {year} {2003})}\BibitemShut
  {NoStop}%
\bibitem [{\citenamefont {Meister}\ \emph {et~al.}(2014)\citenamefont
  {Meister}, \citenamefont {Burger},\ and\ \citenamefont
  {Rauch}}]{Meister_RNS_2014}%
  \BibitemOpen
  \bibfield  {author} {\bibinfo {author} {\bibfnamefont {M.}~\bibnamefont
  {Meister}}, \bibinfo {author} {\bibfnamefont {G.}~\bibnamefont {Burger}}, \
  and\ \bibinfo {author} {\bibfnamefont {W.}~\bibnamefont {Rauch}},\ }\bibfield
   {title} {\enquote {\bibinfo {title} {On the {R}eynolds number sensitivity of
  smoothed particle hydrodynamics},}\ }\href@noop {} {\bibfield  {journal}
  {\bibinfo  {journal} {J. Hydraul. Res.}\ }\textbf {\bibinfo {volume} {52}},\
  \bibinfo {pages} {824--835} (\bibinfo {year} {2014})}\BibitemShut {NoStop}%
\bibitem [{\citenamefont {Lauga}(2007)}]{Lauga_PVF_2007}%
  \BibitemOpen
  \bibfield  {author} {\bibinfo {author} {\bibfnamefont {E.}~\bibnamefont
  {Lauga}},\ }\bibfield  {title} {\enquote {\bibinfo {title} {Propulsion in a
  viscoelastic fluid},}\ }\href@noop {} {\bibfield  {journal} {\bibinfo
  {journal} {Phys. Fluids}\ }\textbf {\bibinfo {volume} {19}},\ \bibinfo
  {pages} {083104} (\bibinfo {year} {2007})}\BibitemShut {NoStop}%
\bibitem [{\citenamefont {Gaffney}\ \emph {et~al.}(2011)\citenamefont
  {Gaffney}, \citenamefont {Gad{\^{e}}lha}, \citenamefont {Smith},
  \citenamefont {Blake},\ and\ \citenamefont
  {Kirkman-Brown}}]{Gaffney_ARFM_2011}%
  \BibitemOpen
  \bibfield  {author} {\bibinfo {author} {\bibfnamefont {E.~A.}\ \bibnamefont
  {Gaffney}}, \bibinfo {author} {\bibfnamefont {H.}~\bibnamefont
  {Gad{\^{e}}lha}}, \bibinfo {author} {\bibfnamefont {D.~J.}\ \bibnamefont
  {Smith}}, \bibinfo {author} {\bibfnamefont {J.~R.}\ \bibnamefont {Blake}}, \
  and\ \bibinfo {author} {\bibfnamefont {J.~C.}\ \bibnamefont
  {Kirkman-Brown}},\ }\bibfield  {title} {\enquote {\bibinfo {title} {Mammalian
  sperm motility: observation and theory},}\ }\href@noop {} {\bibfield
  {journal} {\bibinfo  {journal} {Annu. Rev. Fluid Mech.}\ }\textbf {\bibinfo
  {volume} {43}},\ \bibinfo {pages} {501--528} (\bibinfo {year}
  {2011})}\BibitemShut {NoStop}%
\bibitem [{\citenamefont {Saggiorato}\ \emph {et~al.}(2017)\citenamefont
  {Saggiorato}, \citenamefont {Alvarez}, \citenamefont {Jikeli}, \citenamefont
  {Kaupp}, \citenamefont {Gompper},\ and\ \citenamefont
  {Elgeti}}]{Saggiorato_HSS_2017}%
  \BibitemOpen
  \bibfield  {author} {\bibinfo {author} {\bibfnamefont {G.}~\bibnamefont
  {Saggiorato}}, \bibinfo {author} {\bibfnamefont {L.}~\bibnamefont {Alvarez}},
  \bibinfo {author} {\bibfnamefont {J.~F.}\ \bibnamefont {Jikeli}}, \bibinfo
  {author} {\bibfnamefont {U.~B.}\ \bibnamefont {Kaupp}}, \bibinfo {author}
  {\bibfnamefont {G.}~\bibnamefont {Gompper}}, \ and\ \bibinfo {author}
  {\bibfnamefont {J.}~\bibnamefont {Elgeti}},\ }\bibfield  {title} {\enquote
  {\bibinfo {title} {Human sperm steer with second harmonics of the flagellar
  beat},}\ }\href@noop {} {\bibfield  {journal} {\bibinfo  {journal} {Nat.
  Comm.}\ }\textbf {\bibinfo {volume} {8}},\ \bibinfo {pages} {1415} (\bibinfo
  {year} {2017})}\BibitemShut {NoStop}%
\bibitem [{\citenamefont {Adler}(1946)}]{Adler_SLP_1946}%
  \BibitemOpen
  \bibfield  {author} {\bibinfo {author} {\bibfnamefont {R.}~\bibnamefont
  {Adler}},\ }\bibfield  {title} {\enquote {\bibinfo {title} {A study of
  locking phenomena in oscillators},}\ }\href@noop {} {\bibfield  {journal}
  {\bibinfo  {journal} {Proc. IRE}\ }\textbf {\bibinfo {volume} {34}},\
  \bibinfo {pages} {351--357} (\bibinfo {year} {1946})}\BibitemShut {NoStop}%
\bibitem [{\citenamefont {Niedermayer}\ \emph {et~al.}(2008)\citenamefont
  {Niedermayer}, \citenamefont {Eckhardt},\ and\ \citenamefont
  {Lenz}}]{Niedermayer_SPL_2008}%
  \BibitemOpen
  \bibfield  {author} {\bibinfo {author} {\bibfnamefont {T.}~\bibnamefont
  {Niedermayer}}, \bibinfo {author} {\bibfnamefont {B.}~\bibnamefont
  {Eckhardt}}, \ and\ \bibinfo {author} {\bibfnamefont {P.}~\bibnamefont
  {Lenz}},\ }\bibfield  {title} {\enquote {\bibinfo {title} {Synchronization,
  phase locking, and metachronal wave formation in ciliary chains},}\
  }\href@noop {} {\bibfield  {journal} {\bibinfo  {journal} {Chaos}\ }\textbf
  {\bibinfo {volume} {18}},\ \bibinfo {pages} {037128} (\bibinfo {year}
  {2008})}\BibitemShut {NoStop}%
\bibitem [{\citenamefont {Ishimoto}\ and\ \citenamefont
  {Gaffney}(2018)}]{Ishimoto_HCS_2018}%
  \BibitemOpen
  \bibfield  {author} {\bibinfo {author} {\bibfnamefont {K.}~\bibnamefont
  {Ishimoto}}\ and\ \bibinfo {author} {\bibfnamefont {E.~A.}\ \bibnamefont
  {Gaffney}},\ }\bibfield  {title} {\enquote {\bibinfo {title} {Hydrodynamic
  clustering of human sperm in viscoelastic fluids},}\ }\href@noop {}
  {\bibfield  {journal} {\bibinfo  {journal} {Sci. Rep.}\ }\textbf {\bibinfo
  {volume} {8}},\ \bibinfo {pages} {15600} (\bibinfo {year}
  {2018})}\BibitemShut {NoStop}%
\bibitem [{\citenamefont {Sauzade}\ \emph {et~al.}(2011)\citenamefont
  {Sauzade}, \citenamefont {Elfring},\ and\ \citenamefont
  {Lauga}}]{Sauzade_TSS_2011}%
  \BibitemOpen
  \bibfield  {author} {\bibinfo {author} {\bibfnamefont {M.}~\bibnamefont
  {Sauzade}}, \bibinfo {author} {\bibfnamefont {G.~J.}\ \bibnamefont
  {Elfring}}, \ and\ \bibinfo {author} {\bibfnamefont {E.}~\bibnamefont
  {Lauga}},\ }\bibfield  {title} {\enquote {\bibinfo {title} {Taylor's swimming
  sheet: analysis and improvement of the perturbation series},}\ }\href@noop {}
  {\bibfield  {journal} {\bibinfo  {journal} {Physica D}\ }\textbf {\bibinfo
  {volume} {240}},\ \bibinfo {pages} {1567--1573} (\bibinfo {year}
  {2011})}\BibitemShut {NoStop}%
\bibitem [{\citenamefont {{J{\"u}lich Supercomputing Centre}}(2018)}]{jureca}%
  \BibitemOpen
  \bibfield  {author} {\bibinfo {author} {\bibnamefont {{J{\"u}lich
  Supercomputing Centre}}},\ }\bibfield  {title} {\enquote {\bibinfo {title}
  {{JURECA: Modular supercomputer at J{\"u}lich Supercomputing Centre}},}\
  }\href@noop {} {\bibfield  {journal} {\bibinfo  {journal} {J. Large-Scale
  Res. Facil.}\ }\textbf {\bibinfo {volume} {4}},\ \bibinfo {pages} {A132}
  (\bibinfo {year} {2018})}\BibitemShut {NoStop}%
\end{thebibliography}

%

\end{document}